\documentclass[twocolumn]{aastex701}

\usepackage{amsmath}

\usepackage{siunitx}
\DeclareSIUnit{\jansky}{Jy}
\DeclareSIUnit{\uJy}{\micro\jansky}

\newcommand{\oii}{[\ion{O}{2}]}
\newcommand{\oiii}{[\ion{O}{3}]}
\newcommand{\neiii}{[\ion{Ne}{3}]}
\newcommand{\ciii}{\ion{C}{3}]}
\newcommand{\feii}{\ion{Fe}{2}}
\newcommand{\feiii}{\ion{Fe}{3}}

\begin{document}

\title{An Attenuation Bump at $z=9.3$: Probing Dust Diversity at Cosmic Dawn}

\author[orcid=0000-0002-9415-2296,sname=Rodighiero]{Giulia Rodighiero}
\affiliation{Dipartimento di Fisica e Astronomia “G. Galilei”, Universit\`a di Padova, Vicolo dell'Osservatorio 3, 35122 Padova, Italy}
\affiliation{INAF, Osservatorio Astronomico di Padova, Vicolo dell'Osservatorio 5, 35122 Padova, Italy}
\affiliation{Centro di Ateneo di Studi e Attività spaziali "Giuseppe Colombo" - CISAS, via Venezia, 1 - 35131 - Padova, Italy}
\email[show]{giulia.rodighiero@unipd.it}

\author[orcid=0009-0007-6195-2901]{Santiago Álvarez-Córdoba}
\email{d.alvarez@irya.unam.mx}
\affiliation{Instituto de Radioastronomía y Astrofísica, UNAM, Campus Morelia, A.P. 3-72, Morelia 58089, Mexico}

\author[orcid=0009-0001-8880-944X]{Michele Catone}
\email{michele.catone@studenti.unipd.it}
\affiliation{Dipartimento di Fisica e Astronomia “G. Galilei”, Universit\`a di Padova, Vicolo dell'Osservatorio 3, 35122 Padova, Italy}
\affiliation{INAF, Osservatorio Astronomico di Padova, Vicolo dell'Osservatorio 5, 35122 Padova, Italy}

\author[orcid=0000-0002-0939-9156]{Borja Pérez-Díaz}
\email{borja.perezdiaz@inaf.it}
\affiliation{INAF - Osservatorio Astronomico di Roma, Via di Frascati 33, 00078 Monte Porzio Catone, Italy}

\author[orcid=0000-0002-9400-7312]{Andrea Ferrara}
\email{andrea.ferrara@sns.it }
\affiliation{Scuola Normale Superiore, Piazza dei Cavalieri 7, Pisa I-56126, Italy}
\affiliation{Centro B. Segre, Accademia Nazionale dei Lincei, Via della Lungara, 10, Roma I-00165, Italy}

\author[orcid=0009-0005-6156-4066]{Giorgia Girardi}
\email{giorgia.girardi1@studenti.unipd.it}
\affiliation{Dipartimento di Fisica e Astronomia “G. Galilei”, Universit\`a di Padova, Vicolo dell'Osservatorio 3, 35122 Padova, Italy}
\affiliation{INAF, Osservatorio Astronomico di Padova, Vicolo dell'Osservatorio 5, 35122 Padova, Italy}

\author[orcid=0000-0002-8951-4408]{Lorenzo Napolitano}
\email{lorenzo.napolitano@inaf.it}
=\affiliation{INAF - Osservatorio Astronomico di Roma, Via di Frascati 33, 00078 Monte Porzio Catone, Italy}

\author[orcid=0000-0002-9334-8705]{Paola Santini}
\email{paola.santini@inaf.it}
\affiliation{INAF - Osservatorio Astronomico di Roma, Via di Frascati 33, 00078 Monte Porzio Catone, Italy}

\author[orcid=0000-0003-3596-8794]{Hollis B. Akins}
\altaffiliation{NASA Einstein Fellow}
\email{hollis.akins@gmail.com}
\affiliation{Department of Astrophysical Sciences, Princeton University, 4 Ivy Lane, Princeton, NJ 08544, USA}

\author[orcid=0000-0002-0930-6466]{Caitlin M. Casey}
\email{cmcasey@ucsb.edu}
\affiliation{Department of Physics, University of California, Santa Barbara, Santa Barbara, CA 93106, USA}
\affiliation{Cosmic Dawn Center (DAWN), Denmark}

\author[orcid=0000-0002-3560-8599]{Maximilien Franco}
\email{maximilien.franco@iap.fr}
\affiliation{Institut d’Astrophysique de Paris, UMR 7095, CNRS, Sorbonne Université, 98 bis boulevard Arago, 75014 Paris, France}

\author[orcid=0009-0004-7602-0160]{Pietro Benotto}
\email{pietro.benotto@inaf.it}
\affiliation{Dipartimento di Fisica e Astronomia “G. Galilei”, Universit\`a di Padova, Vicolo dell'Osservatorio 3, 35122 Padova, Italy}
\affiliation{INAF, Osservatorio Astronomico di Padova, Vicolo dell'Osservatorio 5, 35122 Padova, Italy}

\author[orcid=0000-0003-0883-2226]{Rachana Bhatawdekar}
\email{Rachana.Bhatawdekar@esa.int}
\affiliation{European Space Agency (ESAC)}
\affiliation{European Space Agency (ESA), European Space Astronomy Centre (ESAC), Camino Bajo del Castillo s/n, 28692 Villanueva de la Cañada, Madrid, Spain}

\author[orcid=0009-0002-8916-3430]{Alessandro Bianchetti}
\email{alessandro.bianchetti@inaf.it}
\affiliation{INAF—Istituto di Radioastronomia, via Gobetti 101, 40129 Bologna, Italy}

\author[orcid=0000-0003-0492-4924]{Laura Bisigello}
\email{laura.bisigello@inaf.it}
\affiliation{INAF, Osservatorio Astronomico di Padova, Vicolo dell'Osservatorio 5, 35122 Padova, Italy}

\author[orcid=0000-0002-5189-8004]{Daniela Calzetti}
\email{calzetti@astro.umass.edu}
\affiliation{Department of Astronomy, University of Massachusetts Amherst, 710 North Pleasant Street, Amherst, MA 01003, USA}

\author[orcid=0000-0002-6716-4400]{Paolo Cassata}
\email{paolo.cassata@unipd.it}
\affiliation{Dipartimento di Fisica e Astronomia “G. Galilei”, Universit\`a di Padova, Vicolo dell'Osservatorio 3, 35122 Padova, Italy}
\affiliation{INAF, Osservatorio Astronomico di Padova, Vicolo dell'Osservatorio 5, 35122 Padova, Italy}

\author[orcid=0000-0001-9875-8263]{Marco Castellano}
\email{marco.castellano@inaf.it}
\affiliation{INAF - Osservatorio Astronomico di Roma, Via di Frascati 33, 00078 Monte Porzio Catone, Italy}

\author[orcid=0000-0003-1949-7638]{Christopher J. Conselice}
\email{conselice@manchester.ac.uk}
\affiliation{Jodrell Bank Centre for Astrophysics, Alan Turing Building, University of Manchester, Oxford Road, Manchester M13 9PL, UK}

\author[orcid=0009-0007-7842-9930]{Poulomi Dam}
\email{poulomi.dam@studenti.unipd.it}
\affiliation{Dipartimento di Fisica e Astronomia “G. Galilei”, Universit\`a di Padova, Vicolo dell'Osservatorio 3, 35122 Padova, Italy}

\author[orcid=0000-0001-8460-1564]{Pratika Dayal}
\email{pratika.dayal@utoronto.ca}
\affiliation{Canadian Institute for Theoretical Astrophysics, 60 St George St, University of Toronto, Toronto, ON M5S 3H8, Canada}
\affiliation{David A. Dunlap Department of Astronomy and Astrophysics, University of Toronto, 50 St George St, Toronto ON M5S 3H4, Canada}
\affiliation{Department of Physics, 60 St George St, University of Toronto, Toronto, ON M5S 3H8, Canada}

\author[orcid=0000-0002-0200-2857]{Andrea Enia}
\email{andrea.enia@cea.fr}
\affiliation{Universit\'e Paris-Saclay, CEA, IRFU, D\'epartement d'Astrophysique, 91191 Gif-sur-Yvette, France}

\author[orcid=0009-0005-3287-8162]{Gaia Edes Esposito}
\email{gaiaedes.esposito@studenti.unipd.it}
\affiliation{Dipartimento di Fisica e Astronomia “G. Galilei”, Universit\`a di Padova, Vicolo dell'Osservatorio 3, 35122 Padova, Italy}
\affiliation{INAF, Osservatorio Astronomico di Padova, Vicolo dell'Osservatorio 5, 35122 Padova, Italy}

\author[orcid=0000-0002-9382-9832]{Andreas L. Faisst}
\email{afaisst@caltech.edu}
\affiliation{IPAC, California Institute of Technology, 1200 E. California Blvd. Pasadena, CA 91125, USA}

\author[orcid=0000-0003-3248-5666]{Giovanni Gandolfi}
\email{giovanni.gandolfi@inaf.it}
\affiliation{INAF - Osservatorio Astronomico di Roma, Via di Frascati 33, 00078 Monte Porzio Catone, Italy}

\author[orcid=0000-0001-9885-4589]{Steven Gillman}
\email{srigi@space.dtu.dk}
\affiliation{Cosmic Dawn Center (DAWN)}
\affiliation{DTU-Space, Technical University of Denmark, Elektrovej 327, DK2800 Kgs. Lyngby, Denmark}

\author[orcid=0000-0002-5688-0663]{Andrea Grazian}
\email{andrea.grazian@inaf.it}
\affiliation{INAF, Osservatorio Astronomico di Padova, Vicolo dell'Osservatorio 5, 35122 Padova, Italy}

\author[orcid=0000-0002-5836-4056]{Carlotta Gruppioni}
\email{carlotta.gruppioni@inaf.it}
\affiliation{INAF -- Osservatorio di Astrofisica e Scienza dello Spazio di Bologna,
Via P. Gobetti 93/3, 40129 Bologna, Italy}

\author[orcid=0000-0002-3301-3321]{Michaela Hirschmann}
\email{michaela.hirschmann@epfl.ch}
\affiliation{Institute of Physics, Lab for galaxy evolution, EPFL, Observatory of Sauverny, Chemin Pegasi 51, 1290 Versoix, Switzerland}

\author[orcid=0000-0001-8386-3546,sname=Iani]{Edoardo Iani}
\email[]{} 
\affiliation{Institute of Science and Technology Austria (ISTA), Am Campus 1, 3400 Klosterneuburg, Austria}

\author[orcid=0000-0001-9840-4959]{Kohei Inayoshi}
\email{inayoshi@pku.edu.cn}
\affiliation{Kavli Institute for Astronomy and Astrophysics, Peking University, Beijing 100871, China}

\author[orcid=0000-0001-9187-3605]{Jeyhan S. Kartaltepe}
\email{jeyhan@astro.rit.edu}
\affiliation{Laboratory for Multiwavelength Astrophysics, School of Physics and Astronomy, Rochester Institute of Technology, 84 Lomb Memorial Drive, Rochester, NY 14623, USA}

\author[orcid=0000-0002-4872-2294]{Georgios Magdis}
\email{geoma@dtk.dk}
\affiliation{Cosmic Dawn Center (DAWN)}
\affiliation{DTU-Space, Technical University of Denmark, Elektrovej 327, DK2800 Kgs. Lyngby, Denmark}

\author[orcid=0000-0002-9889-4238]{Alessandro Marconi}
\email{alessandro.marconi@unifi.it}
\affiliation{Dipartimento di Fisica e Astronomia,
Universit\`a degli Studi di Firenze,
Via G. Sansone 1, 50019 Sesto Fiorentino, Italy}
\affiliation{INAF -- Osservatorio Astrofisico di Arcetri,
Largo Enrico Fermi 5, 50125 Firenze, Italy}

\author[orcid=0000-0001-9189-7818]{Crystal L. Martin}
\email{cmartin@ucsb.edu}
\affiliation{Department of Physics, University of California Santa Barbara, Santa Barbara, CA 93106, USA}

\author[orcid=0000-0002-2281-2785]{Bo Milvang-Jensen}
\email{milvang@astro.ku.dk}
\affiliation{Cosmic Dawn Center (DAWN)}
\affiliation{Niels Bohr Institute, University of Copenhagen, Jagtvej 128, 2200 Copenhagen N, Denmark}

\author{Annagrazia Puglisi}
\email[]{} 
\affiliation{School of Physics and Astronomy, University of Southampton, Southampton SO17 1BJ, UK}

\author[orcid=0000-0002-7093-7355]{Alvio Renzini}
\email{alvio.renzini@inaf.it} 
\affiliation{INAF, Osservatorio Astronomico di Padova, Vicolo dell'Osservatorio 5, 35122 Padova, Italy}

\author[orcid=0000-0003-0427-8387]{R. Michael Rich}
\email{rmr@astro.ucla.edu}
\affiliation{Department of Physics and Astronomy, UCLA, Los Angeles, CA, 90095-1547}

\author[orcid=0000-0002-5104-8245]{Pierluigi Rinaldi}
\email[]{} 
\affiliation{Department of Astronomy, The University of Texas at Austin, Austin, TX 78712, USA}
\affiliation{Cosmic Frontier Center, The University of Texas at Austin, Austin, TX 78712, USA}

\author[orcid=0000-0002-4271-0364]{Brant E. Robertson}
\email{brant@ucsc.edu}
\affiliation{Department of Astronomy and Astrophysics, University of California, Santa Cruz, 1156 High Street, Santa Cruz, CA 95064, USA}

\author[orcid=0000-0002-0517-7943]{Stephen Serjeant}
\email{stephen.serjeant@open.ac.uk}
\affiliation{School of Physical Sciences, The Open University, Walton Hall, Milton Keynes, MK7 6AA, UK}

\author[orcid=0000-0003-4702-7561]{Irene Shivaei}
\email{ishivaei@cab.inta-csic.es}
\affiliation{Centro de Astrobiolog\'{i}a (CAB), CSIC-INTA, Carretera de Ajalvir km 4, Torrej\'{o}n de Ardoz, 28850, Madrid, Spain}

\author[orcid=0000-0003-3631-7176]{Sune Toft}
\email{sune@nbi.ku.dk}
\affiliation{Cosmic Dawn Center (DAWN)}
\affiliation{Niels Bohr Institute, University of Copenhagen, Jagtvej 128, 2200 Copenhagen N, Denmark}

\author[orcid=0000-0003-4032-0853]{Amelia Vietri}
\email[]{} 
\affiliation{Dipartimento di Fisica e Astronomia “G. Galilei”, Universit\`a di Padova, Vicolo dell'Osservatorio 3, 35122 Padova, Italy}
\affiliation{INAF, Osservatorio Astronomico di Padova, Vicolo dell'Osservatorio 5, 35122 Padova, Italy}

\author[orcid=0000-0003-0980-1499]{Benedetta Vulcani}
\email{benedetta.vulcani@inaf.it} 
\affiliation{INAF, Osservatorio Astronomico di Padova, Vicolo dell'Osservatorio 5, 35122 Padova, Italy}

\author[orcid=0000-0003-4793-7880]{Fabian Walter}
\email{walter@mpia.de}
\affiliation{Max Planck Institut f\"ur Astronomie, K\"onigstuhl 17, D-69117 Heidelberg, Germany}

\author[orcid=0000-0003-3903-6935]{Stephen M.~Wilkins}
\email{s.wilkins@sussex.ac.uk} 
\affiliation{Astronomy Centre, University of Sussex, Falmer, Brighton BN1 9QH, UK}
\affiliation{Institute of Space Sciences and Astronomy, University of Malta, Msida MSD 2080, Malta}

\author[0000-0002-7051-1100]{Jorge A. Zavala}\email{jzavala@umass.edu}
\affiliation{Department of Astronomy, University of Massachusetts Amherst, 710 North Pleasant Street, Amherst, MA 01003, USA}


\begin{abstract}
We report the identification of a UV attenuation bump in the Black Pearl, a galaxy at $z_{\rm spec}=9.3335$ observed with JWST/NIRSpec as part of the ZENITH and EMBER surveys. This is the highest-redshift UV bump reported to date, probing dust properties only $\sim500\,\mathrm{Myr}$ after the Big Bang. The feature is directly detected as a broad depression in the rest-frame UV continuum and is well described by a local Drude profile, with a central wavelength $\lambda_{\max}=2132\pm17\,\AA$, an FWHM $=413\pm54\,\AA$, and a peak attenuation $A_{\max}=0.21\pm0.02\,\mathrm{mag}$. Joint spectrophotometric modeling with a flexible attenuation law independently recovers a pronounced UV bump without imposing its presence a priori, regardless of the adopted star-formation history. The modeling indicates a relatively low-mass, actively star-forming galaxy, with $\log(M_\star/M_\odot)=8.35^{+0.09}_{-0.08}$, a star-formation rate of $2.36^{+0.54}_{-0.42}\,M_\odot\,{\rm yr}^{-1}$, and $A_V=0.34^{+0.12}_{-0.11}\,\mathrm{mag}$. Rest-frame optical emission-line diagnostics indicate a hard ionizing spectrum and allow for a moderate AGN contribution, but provide no evidence for a powerful AGN capable of dominating the observed UV continuum. The bump is blueshifted by $43\pm17\,\AA$ ($\simeq2.5\sigma$) relative to the canonical Milky Way $2175\,\AA$ feature and by $127\pm17\,\AA$ relative to the few UV bumps currently known at $z\sim7$, while its width is consistent with the broad distribution measured along Milky Way sightlines. The displaced central wavelength therefore provides the clearest evidence for a nonstandard UV-bump profile at $z=9.3335$, pointing to substantial diversity in the properties of UV-bump carriers at early cosmic times, with both PAHs and small graphite grains remaining plausible candidates.

\end{abstract}


\section{Introduction}
\label{intro}

The rapid enrichment of galaxies with metals and dust during the
first billion years of cosmic history is emerging as one of the major
discoveries enabled by the \textit{James Webb} Space Telescope \citep[JWST,][]{Gardner2023}. Recent spectroscopic observations have revealed
galaxies with intense nebular emission, hard ionizing radiation fields,
and substantial metal abundances already in place during the epoch of
reionization and beyond, only a few hundred million years after the
Big Bang
\citep[e.g.,][]{Curti2023,Fujimoto2023,Sanders2024,Curti2025,Faisst2026}.
At the same time, Atacama Large Millimeter Array observations have demonstrated that significant
dust reservoirs and obscured star formation are already present in
some galaxies at $z\sim7\mbox{--}8$
\citep[e.g.,][]{Dayal2022,Inami2022,Sommovigo2022,Ferrara2022,Algera2023,Bakx2025}, as well as in 
host galaxies of luminous quasars at similar epochs \citep[e.g.,][]{Walter2022}.
Evidence for substantial dust enrichment is now emerging at even earlier
epochs: the spectroscopically confirmed galaxy EGS-z11-R0 at $z=11.45$
exhibits a remarkably red UV continuum and significant dust attenuation,
demonstrating that dust-enriched systems can already be in place within
the first $\sim400\,\mathrm{Myr}$ of cosmic history in a massive $\log_{10}(M_\star/M_\odot)\sim9.2$ galaxy \citetext{\citealp{Rodighiero2026}, but see also
\citealp{Donnan2025} and \citealp{Mitsuhashi2026}}.
Together, these results are transforming our understanding of the
time scales of metal production, dust growth, and interstellar medium (ISM) processing in the early Universe and their interplay \citep[e.g.,][]{Markov2025,Narayanan2025,Burgarella25}.

In this context, the ultraviolet (UV) extinction bump centered approximately at
 $2175\,\AA$ represents a particularly powerful diagnostic
of dust composition and grain evolution
\citep[e.g.,][]{FitzpatrickMassa1986,Cardelli1989,Draine2003}.
The feature is the strongest  structure in 
the interstellar extinction curve in the Milky Way and is also observed
in nearby galaxies and in absorption systems over a broad range of
redshifts \citep[e.g.,][]{Gordon2003,Noll2009,Gordon2024}.
The UV bump is widely attributed to small carbonaceous grains or molecules, although the precise identity and physical properties of its carriers remain uncertain. The proposed candidates include small graphite grains, polycyclic aromatic hydrocarbons (PAHs), amorphous carbonaceous materials, and related aromatic carbonaceous particles \citep[e.g.,][]{LiDraine2001,Draine2003,Bradley2005,Papoular2009,Lin2025}. 
The formation, survival, and
processing of these carriers depend on the physical conditions of the
ISM, making the presence and detailed profile of the UV bump a
potentially sensitive probe of dust evolution.

Before JWST, studies of the UV bump beyond the local Universe were largely limited to galaxies at intermediate redshifts \citep[e.g.,][]{Noll2009,Shivaei2022}. In particular, \citep{Kashino2021} reported evidence for the 2175\,\AA\ attenuation feature from a large statistical sample of more than 500 galaxies at $z\sim2$, demonstrating that the bump was already a widespread component of dust attenuation at these epochs. JWST spectroscopy has since extended direct investigations of the feature into the epoch of reionization.
A landmark result was the spectroscopic identification of a pronounced UV bump
in JADES-GS-z6-0 at $z=6.71$ \citep{Witstok2023}, demonstrating that
a dusty population capable of producing a structured UV attenuation
feature was already present only $\sim800\,\mathrm{Myr}$ after the Big Bang.
Remarkably, the feature peaks at
$\lambda_{\rm max}\simeq2263\,\AA$, at a longer wavelength than the
canonical Milky Way bump, already suggesting possible differences in
the underlying grain population \citep{Witstok2023,Lin2025}.

Subsequent JWST observations have strengthened the evidence
that UV bumps were already present during the reionization era. A
strong $2175\,\AA$ feature has been identified in the $z=7.11$ galaxy
GNWY-7379420231 \citep{Ormerod2025}, while larger spectroscopic samples
have revealed additional candidates extending to $z\simeq7.5$ and
suggest an overall evolution of dust attenuation properties with
redshift \citep{Markov2025}. In particular, the average decrease in
UV-bump strength toward earlier cosmic times has been interpreted as
evidence for dust populations increasingly dominated by large grains
recently condensed in supernova ejecta and subject to comparatively
little processing in the ISM \citep{Markov2025}. The existence of
individual galaxies with pronounced bumps at these epochs therefore
points to substantial diversity in the pace of dust enrichment and
processing during the first billion years.

The unusual UV-bump profiles observed at high redshift may reflect differences in the properties of their carbonaceous carriers. Recent modeling of small graphitic grains and PAH populations has highlighted the sensitivity of the bump position and width to the nature and mixture of these carriers \citep{Li2024,Lin2025}.
Observational studies at lower redshift have also explored a connection between PAH abundance and UV-bump strength \citep{Shivaei2022,Battisti2025}, although the relationship does not uniquely identify the carriers.


The diversity of the observed UV-bump profiles may provide 
important information on the physical properties of early dust.
In the Milky Way, the central wavelength of the bump is remarkably
stable, whereas its width and strength exhibit substantial variations
among different sightlines \citep{FitzpatrickMassa1986}. At high
redshift, the still limited number of spectroscopic detections already
appears to span a broader range of bump properties, raising the
possibility that the carriers and/or their processing state differ
among galaxies. In this perspective, the central wavelength, width,
and strength of the UV bump become complementary observables that can
help constrain the composition, size distribution, and local
processing of the underlying grain population.

Such constraints are particularly valuable at Cosmic Dawn, where the
limited time available for dust production and processing places
strong restrictions on the origin of the grain population. They are
especially valuable at redshifts, where direct constraints on dust
properties from its thermal emission remain extremely challenging,
making rest-frame UV attenuation features one of the few available
probes of the composition and processing state of the grain population.
Recent studies suggest that dust attenuation curves evolve
systematically with redshift, becoming on average flatter and
displaying weaker UV bumps toward earlier cosmic times
\citep{Shivaei2020,Shivaei2025,Markov2025,Rodighiero2026}. This behavior has been interpreted as evidence
for early dust populations dominated by relatively large grains
freshly condensed in supernova ejecta, followed by the progressive
build-up of smaller grains through growth and processing in the ISM
\citep{Markov2025,Narayanan2026} together with variations in the
star--dust geometry \citep{Matsumoto2026}. The detection of pronounced UV
bumps in individual galaxies during the reionization era therefore
suggests that dust evolution may proceed rapidly, but not necessarily
uniformly, across the early galaxy population.

The UV bump also has important implications for modeling 
high-redshift galaxy spectral energy distributions (SEDs). The presence of
a structured attenuation feature can modify the inferred UV continuum slope and normalization
and, consequently, the physical properties derived from SED fitting
\citep[e.g.,][]{Shivaei2022}. Flexible attenuation prescriptions are
therefore particularly valuable at high redshift, where the shape of
the attenuation curve cannot be assumed a priori. 
At the same time, broad spectral structures (including blended
\feii, \feiii{} emission) producing a UV pseudocontinuum 
may affect the apparent UV continuum
\citep{VestergaardWilkes2001}. This might be caused by the presence of an AGN, although shocks can significantly enhance the Fe content of the ISM \citep{Mendez2024,Scheffler2026} producing bright \feii{} and \feiii{} emission as cooling \citep[e.g.][]{Bouvis2025}.
An AGN origin can, however, be
independently assessed through characteristic emission-line
diagnostics.  Spectroscopic identification of the
feature, together with independent constraints on the nature of the
ionizing source, is therefore essential for establishing a robust
dust interpretation.

In this paper, we report the identification of a UV attenuation bump in the Black Pearl, a galaxy at $z_{\rm spec}=9.3335$, corresponding to a cosmic age of only $\sim 500\,\mathrm{Myr}$ after the Big Bang and representing the highest-redshift detection of this feature reported to date.
We characterize the feature directly from the JWST/NIRSpec
spectrum and independently through joint spectrophotometric modeling
with a flexible attenuation law. We further use rest-frame optical
emission-line diagnostics to investigate the nature of the ionizing
source and assess possible alternative interpretations of the UV
continuum, including an AGN-related \feii/\feiii{} pseudocontinuum.

Beyond establishing the presence of the feature, we investigate its
central wavelength, width, and strength and compare its profile with
the canonical Milky Way bump and with the few UV bumps currently known
during the reionization era. We use these measurements to explore
whether the diversity of UV-bump profiles can provide new constraints
on the composition and processing state of dust at Cosmic Dawn.

Throughout this work, we adopt a standard flat $\Lambda$CDM cosmology
with $H_0=70\,\mathrm{km}\,\mathrm{s}^{-1}\,\mathrm{Mpc}^{-1}$, $\Omega_{\rm M}=0.3$, 
$\Omega_{\Lambda}=0.7$ and a \citet{Kroupa2002} initial mass function (IMF).

\section{Target identification and observations}
\label{sec:data}

The galaxy analyzed in this work was selected as a primary target of the ZENITH program (JWST GO-7417; PI: C. Casey), a NIRSpec spectroscopic survey designed to confirm and characterize intrinsically luminous galaxy candidates at $z>10$ selected from COSMOS-Web. ZENITH was organized in conjunction with the Early Massive Black Hole Evolution \& Reddening (EMBER; JWST GO-7076; PI: H. Akins) program, which targets primarily Little Red Dots (LRDs) at $z\sim3\mbox{--}7$. The two programs share a number of targets and coordinated NIRSpec configurations. The source studied here is a primary ZENITH high-redshift target observed on an EMBER NIRSpec mask.

The JWST/NIRSpec observations were obtained as part of the ZENITH program (GO-7417), using the multi-object spectroscopy mode with the Micro-Shutter Assembly (MSA). The low-resolution PRISM/CLEAR configuration provides continuous wavelength coverage from approximately $0.6$ to $5.3\,\micron$, with a wavelength-dependent resolving power of $R \sim 30\mbox{--}300$ \citep{Jakobsen2022,Ferruit2022}.
The observations employed three-shutter slitlets and a nodding pattern along the slitlets to facilitate background subtraction.
The NIRSpec data were reduced starting from the uncalibrated exposures using the JWST Calibration Pipeline \citep{Bushouse2023}. In that reduction, the standard detector-level calibrations were supplemented by an iterative correction for the thermal picture-frame pattern and $1/f$ noise, with spectral traces masked during background estimation. Subsequent processing included flat-fielding, wavelength and flux calibration and combination of the individual exposures into rectified two-dimensional spectra. The final one-dimensional spectrum was obtained using an optimal extraction procedure following \citet{Horne1986}.

The NIRSpec spectroscopy is complemented by broadband photometry spanning the observed optical to near-infrared wavelength range (Table\,\ref{tab:photometry}). The Hubble Space Telescope (HST) ACS imaging is drawn from the COSMOS and CANDELS surveys \citep{Scoville2007,Koekemoer2007,Koekemoer2011,Grogin2011}. The JWST/NIRCam observations comprise the F115W, F150W, F277W, and F444W bands from COSMOS-Web  \citep{Casey2023}, together with F200W and F356W imaging from COSMOS-3D (JWST GO-5893; PI: K. Kakiichi). The imaging reduction broadly follows the procedures described by \citet{Franco2026}. Photometric measurements were extracted from an internally developed multisurvey catalog incorporating the available HST and JWST imaging, using an aperture-based approach similar to that adopted for the COSMOS2025 catalog \citep{Shuntov2025}. The resulting flux densities and uncertainties used in this work are reported in Table\,\ref{tab:photometry}.


\begin{table}
\centering
\caption{Broadband photometry of the Black Pearl. Flux densities
and $1\sigma$ uncertainties are reported in $\si{\nano\jansky}$.}
\label{tab:photometry}
\begin{tabular}{llcc}
\hline
\hline
Filter & Instrument & $\lambda_{\rm eff}$ [$\micron$] &
$F_\nu \pm \sigma_{F_\nu}$ [$\si{\nano\jansky}$] \\
\hline
F606W & HST/ACS     & 0.5960 & $-9.38 \pm 8.04$ \\
F814W & HST/ACS     & 0.8073 & $4.23 \pm 11.79$ \\
F115W & JWST/NIRCam & 1.1570 & $17.02 \pm 9.53$ \\
F150W & JWST/NIRCam & 1.5040 & $78.24 \pm 12.40$ \\
F200W & JWST/NIRCam & 1.9930 & $77.24 \pm 5.89$ \\
F277W & JWST/NIRCam & 2.7690 & $78.42 \pm 7.38$ \\
F356W & JWST/NIRCam & 3.5770 & $78.69 \pm 5.02$ \\
F444W & JWST/NIRCam & 4.4160 & $111.30 \pm 9.44$ \\
\hline
\end{tabular}
\end{table}

In the context of the present study, we visually inspected the NIRSpec
spectra of high-redshift sources targeted by the coordinated ZENITH/EMBER observations, focusing in
particular on objects at $z>8$ for which the rest-frame wavelength
region encompassing the $\sim2175\,\AA$ UV attenuation feature is
covered by the observations. The inspection was aimed at identifying
broad continuum structures potentially associated with dust attenuation,
rather than selecting sources on the basis of a pre-existing UV-bump
signature.

This search identified a particularly interesting source, hereafter referred to as the Black Pearl, at $\mathrm{R.A. = 10^{h}01^{m}09^{s}.35}$ and $\mathrm{Dec. = +02^{\circ}03'19''.3}$ (J2000). The galaxy is identified as \texttt{ember\_cosmos\_p8\_251119} in the survey catalog.
The target was observed with JWST/NIRSpec using both the PRISM/CLEAR and
G395M/F290LP configurations, with total integration times of 1.5\,hr and
55\,min, respectively. The PRISM spectrum provides the broad wavelength
coverage required to characterize the overall spectral energy
distribution and rest-frame UV continuum, while the G395M observations
provide complementary higher-resolution information over the
corresponding wavelength range.

The NIRSpec spectrum reveals a well-detected UV continuum together with
a prominent broad depression around the wavelength expected for the UV
attenuation bump. 
We determine a spectroscopic redshift of
$z_{\rm spec}=9.3335\pm0.0007$ from the line centroids of
H$\beta$ and the \oiii$\lambda\lambda4959,5007$ doublet
in the NIRSpec/PRISM spectrum (see Sect.\,\ref{sec:emission_lines}). At this redshift, the observed
broad feature lies in the rest-frame UV, making this source an
exceptional candidate for investigating the presence of chemically
evolved dust at very early cosmic times.

\begin{figure}
\includegraphics[width=\columnwidth]{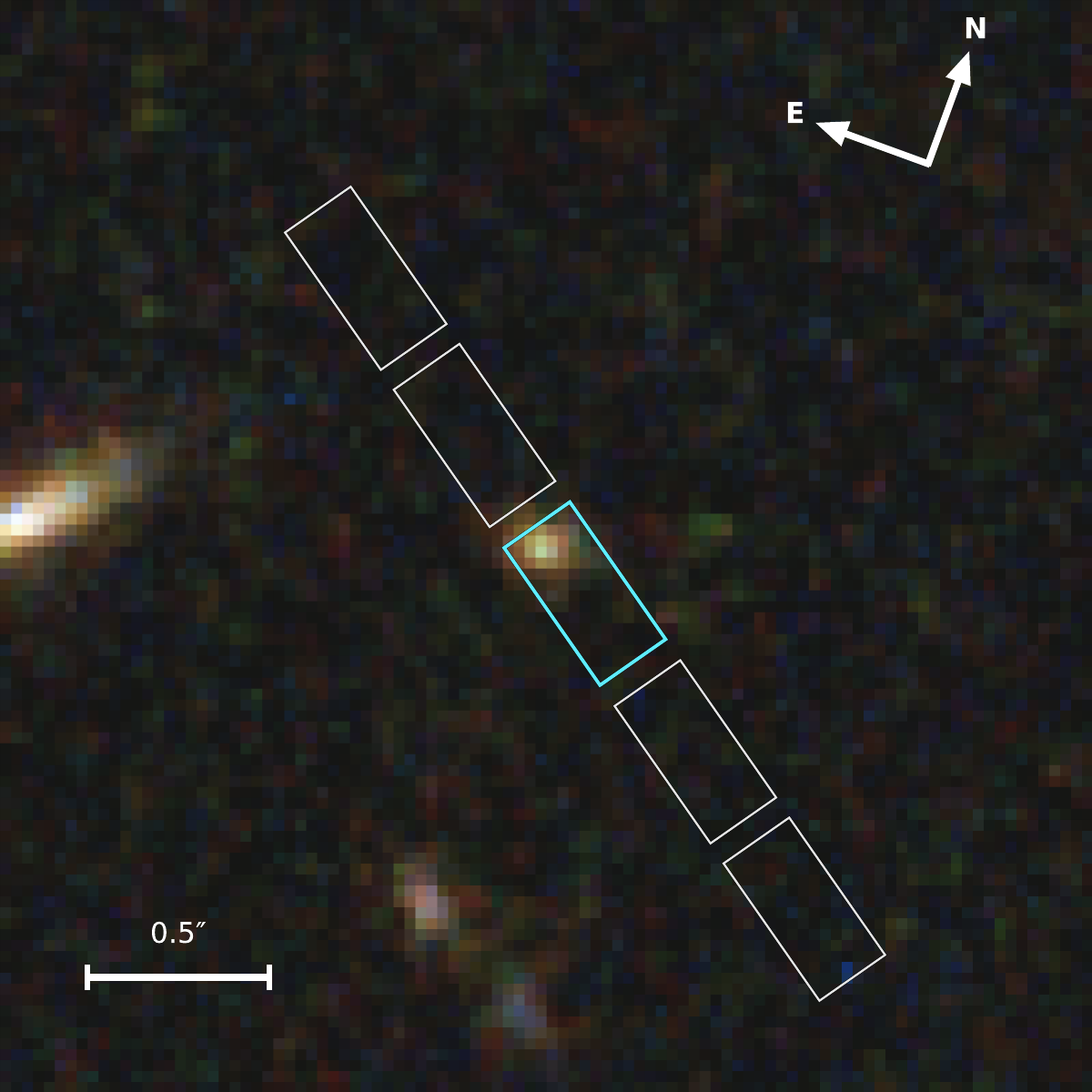}
\caption{
JWST/NIRCam color cutout obtained by combining the six NIRCam
filters, centered on the Black Pearl. The overlaid cyan rectangles show the footprint of the NIRSpec MSA associated with
the spectroscopic observations. The  cutout spans 3 arcsec on a side.
}
\label{fig:target_cutout}
\end{figure}

\begin{figure*}
\includegraphics[width=18cm]{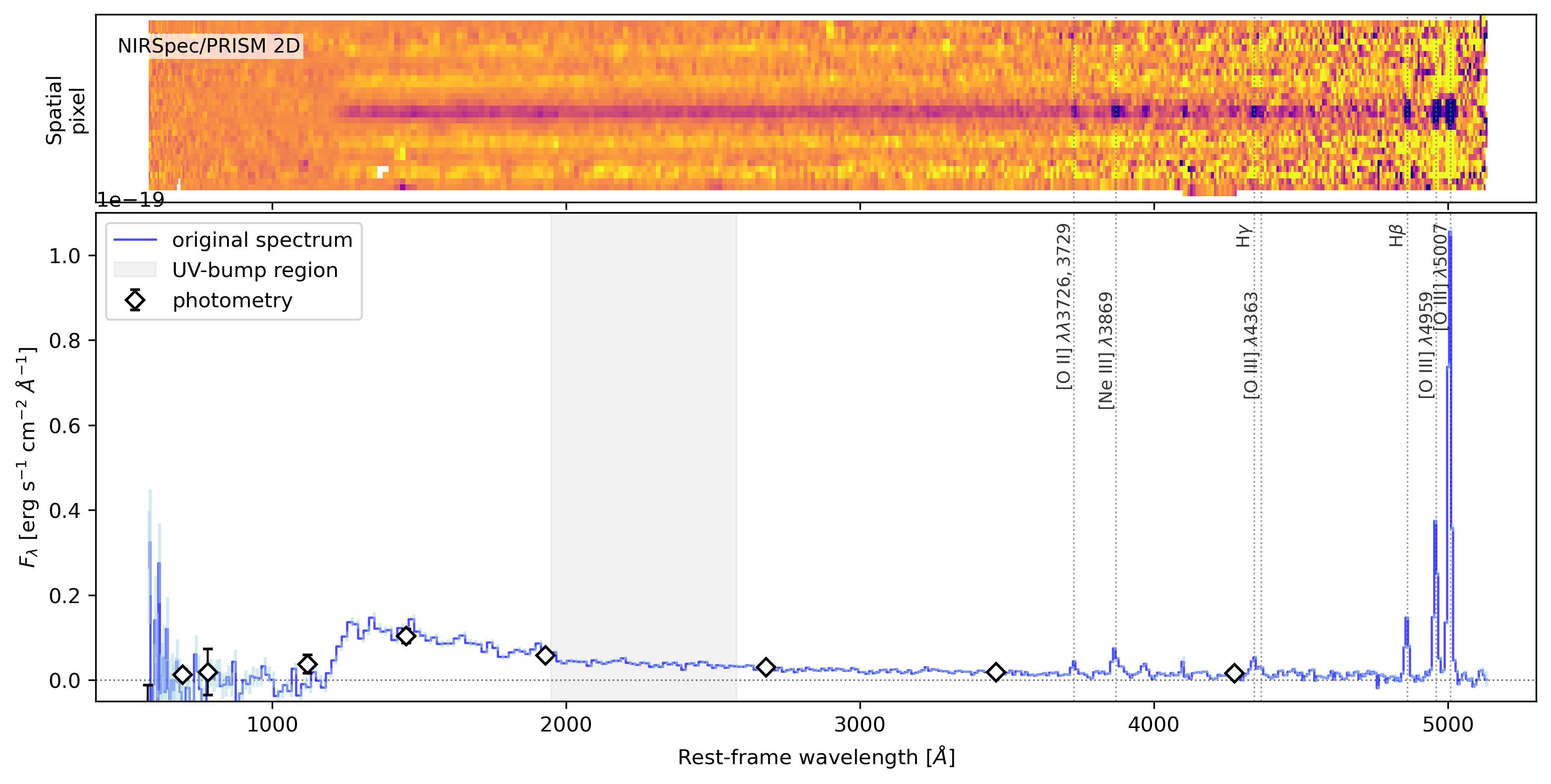}
\caption{
Rest-frame JWST/NIRSpec PRISM spectrum of the Black Pearl at $z_{\rm spec} = 9.3335$, in units of $10^{-19}\,{\rm erg\,s^{-1}\,cm^{-2}\,\AA^{-1}}$.
The top panel shows the two-dimensional NIRSpec/PRISM spectrum, while the
bottom panel shows the extracted one-dimensional spectrum. broadband
photometric measurements are shown as open diamonds. The shaded region
highlights the wavelength interval encompassing the broad UV attenuation
feature. Vertical dotted lines mark the rest-frame wavelengths of the main
emission lines detected at integrated ${\rm S/N}\geq3$ and reported in
Table~\ref{tab:lines}.
}
\label{fig:spectrum}
\end{figure*}

\begin{figure*}
\includegraphics[width=18cm]{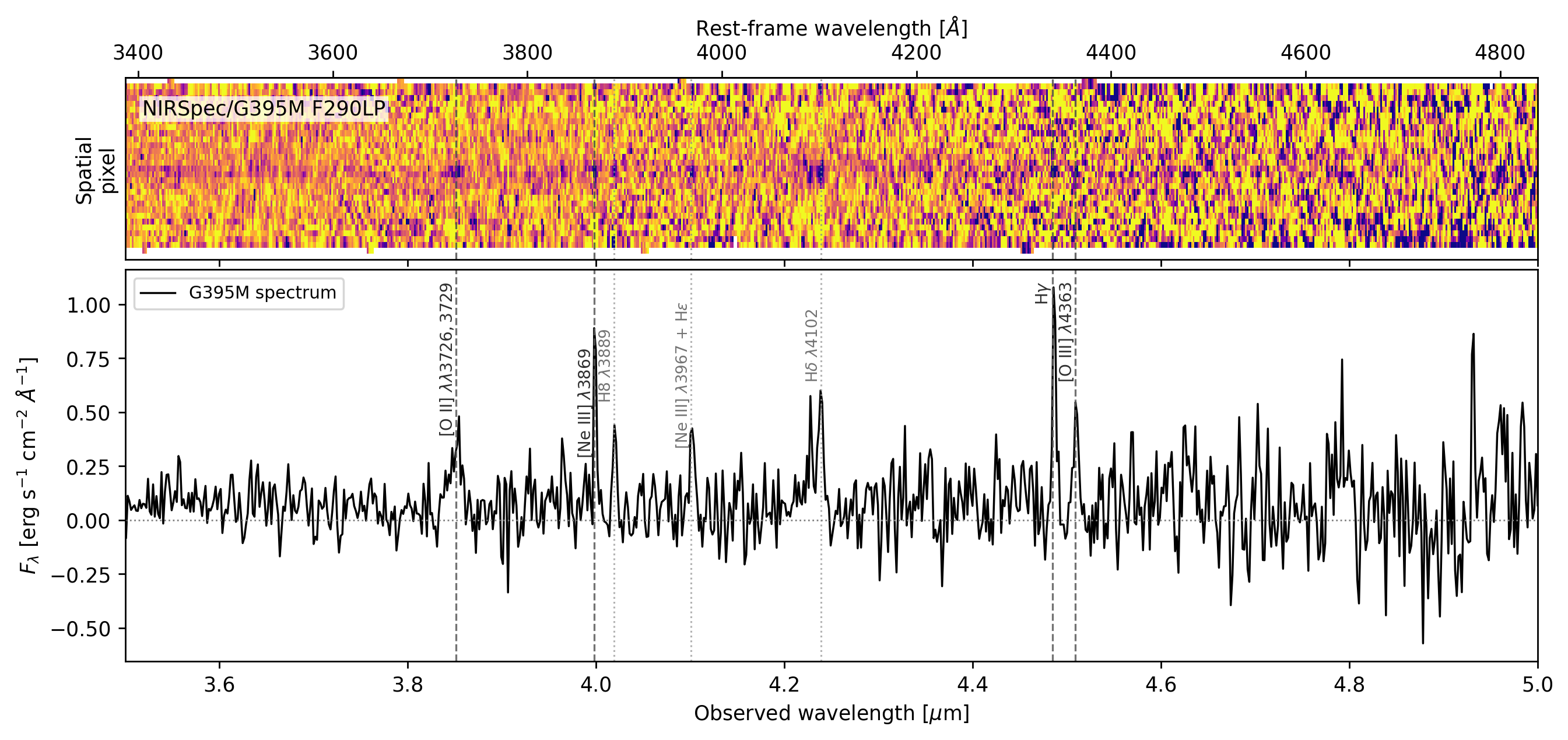}
\caption{JWST/NIRSpec G395M/F290LP spectroscopy of the 
$z_{\rm spec}=9.3335$ galaxy, in units of $10^{-19}\,{\rm erg\,s^{-1}\,cm^{-2}\,\AA^{-1}}$. The top panel shows the two-dimensional 
spectrum, while the bottom panel shows the extracted one-dimensional 
spectrum. Dashed vertical lines mark the expected wavelengths of the 
emission features measured ($\mathrm{S/N}>3$) in the G395M spectrum: 
\oii$\lambda\lambda$3726,3729, \neiii$\lambda3869$, 
H$\gamma$, and \oiii$\lambda4363$. Lighter dotted lines indicate 
additional candidate features  ($\mathrm{S/N}<3$) at the expected positions of 
H8 $\lambda3889$, the \neiii$\lambda3967$+H$\epsilon$ blend, 
and H$\delta$ $\lambda4102$. These additional features are shown 
for reference but are not included among the formally measured ($\mathrm{S/N}>3$)
emission lines reported in Table\,\ref{tab:lines}.}\label{fig:spectrum_grism}
\end{figure*}

In the following, we use the JWST/NIRSpec spectra together with the
available broadband photometry to characterize the source. We first
examine the spectral feature directly and test whether its shape is
consistent with a UV attenuation bump (Sect.\,\ref{sec:spectral}), before
performing the joint spectrophotometric modeling described in
Sect.\,\ref{sed_fitting}.

Figure\,\ref{fig:target_cutout} shows the NIRCam RGB cutout centered on
on the Black Pearl, together with the footprint of the
NIRSpec MSA shutters used for the spectroscopic observations.

\section{Spectral analysis}\label{sec:spectral}

\begin{figure*}
\includegraphics[width=18cm]{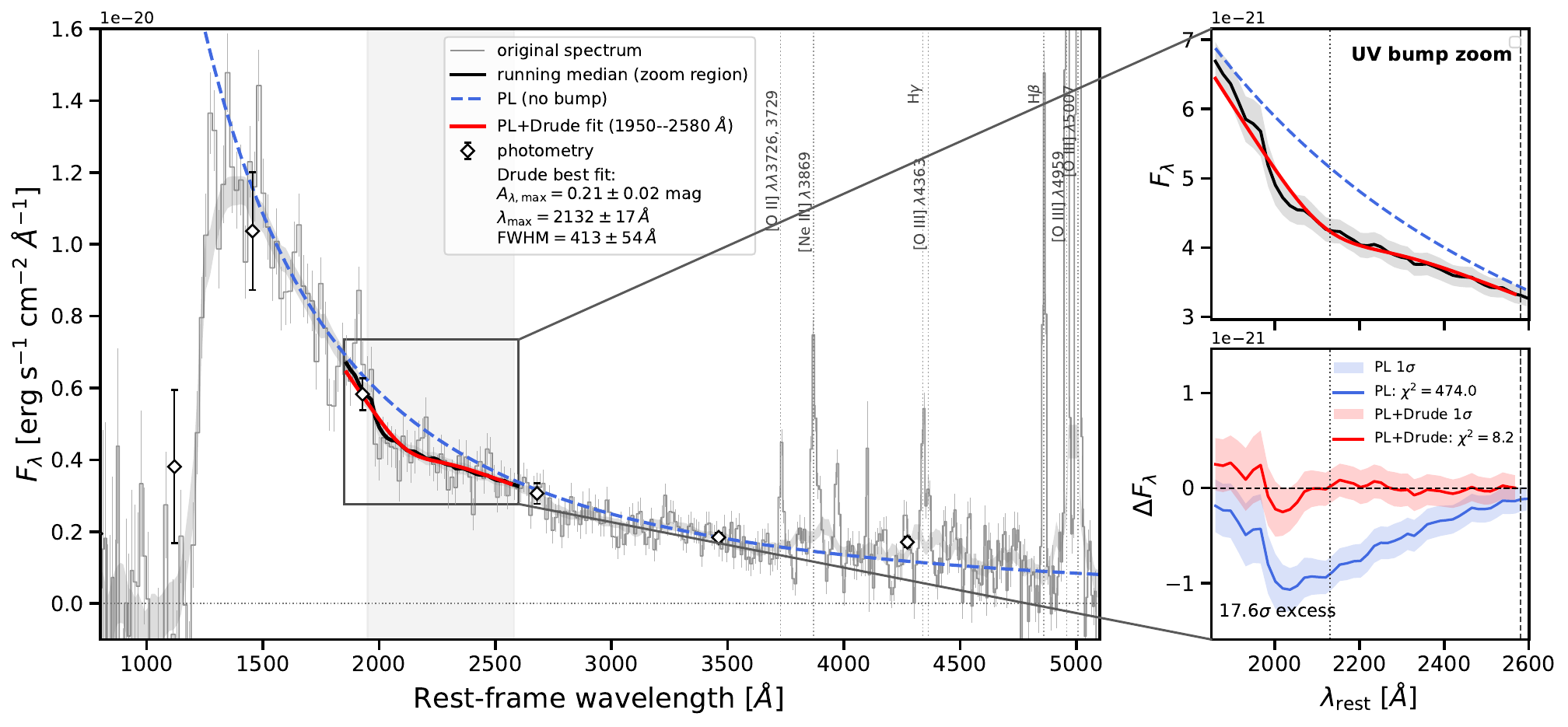}
\caption{
Direct spectral characterization and statistical significance of the broad UV
attenuation feature. 
In the left panel, the gray histogram shows the original NIRSpec/PRISM spectrum
at its native spectral sampling, while the black curve
shows the running-median spectrum, with the surrounding
gray shaded region indicating its $1\sigma$ uncertainty
estimated from Monte Carlo perturbations of the original
spectral pixels. The vertical gray shaded band marks the
wavelength interval adopted for the local UV-bump fit.
The broadband photometry is reported with open diamonds.
The blue dashed curve is the local power-law continuum, while the red curve shows the
power-law plus best-fitting Drude profile over the interval used for the local
bump fit. The fit gives $A_{\rm max}=0.21\pm0.02\,\mathrm{mag}$,
$\lambda_{\rm max}=2132\pm17\,\AA$, and ${\rm FWHM}=413\pm54\,\AA$.
Vertical dotted lines indicate the principal identified emission lines.
The upper-right panel enlarges the UV-bump region, showing the smoothed
spectrum and its $1\,\sigma$ uncertainty together with the power-law (PL) and
PL + Drude models. The lower-right panel shows the residuals relative to the
power-law continuum (blue) and PL+Drude model (red), with the corresponding
$1\,\sigma$ uncertainties. 
}
\label{fig:drude}
\end{figure*}

The rest-frame UV-to-optical PRISM spectrum of the galaxy is shown 
in Fig.\,\ref{fig:spectrum}, while the complementary medium-resolution G395M spectrum 
is presented in Fig.\,\ref{fig:spectrum_grism}.
Before proceeding with the analysis, we performed a slit-loss correction by matching the NIRSpec spectrum to the NIRCam photometry. To do so, we compared the observed broadband photometry with synthetic photometry computed by convolving the spectrum with the NIRCam filter transmission curves (F150W, F200W, F277W, F356W, and F444W, while F115W was excluded due to its comparatively large flux uncertainty). For each filter, synthetic flux densities were obtained as the transmission-weighted integral of the spectrum, with errors propagated accordingly. The wavelength-dependent ratio between the photometric and synthetic fluxes was then fit with a linear function of wavelength, weighted by the combined photometric and spectroscopic uncertainties. The resulting best-fit relation was applied as a multiplicative correction to the full spectrum. The correction increases the flux by $\sim 20\%$ on average (mean factor $\simeq 1.21$), ranging from $\sim 1.02$ at the blue end of the spectrum ($0.6\,\micron$) to $\sim 1.36$ at the red end ($5.3\,\micron$). Uncertainties on the corrected flux were propagated from both the spectroscopic errors and the uncertainties of the fitted correction coefficients.
On the corrected spectrum, the continuum is well detected over a broad wavelength range  and exhibits a relatively blue UV slope, with 
$\beta=-2.22\pm0.08$ from a power-law fit over the rest-frame 
$1350\mbox{--}2600\,\AA$ interval ($\chi^2_{\rm red}=1.2$). We note, however, 
that this wavelength range encompasses the UV-bump region, and the 
derived slope may therefore be affected by the broad attenuation 
feature discussed below. 
With an observed UV absolute magnitude of $M_{\rm UV}=-20.619\pm0.012$, the galaxy ranks among the  most UV-luminous spectroscopically confirmed sources currently known at $z>9$.

The broad continuum depression around the expected UV-bump region is
already apparent in the NIRSpec spectrum. We characterize this feature
directly from the spectrum in Sect.~\ref{identification}, while an independent assessment based on joint spectrophotometric modeling with a flexible attenuation
law is presented in Sect.~\ref{sed_fitting}.

\subsection{Identification of the broad UV absorption feature}
\label{identification}

We first investigated whether the broad depression has the characteristic
shape expected for a UV attenuation bump. The feature is directly visible in
the NIRSpec spectrum, and the analysis below characterizes its profile and
statistical significance independently of the full spectrophotometric
modeling. To this end, we modeled the
local UV continuum with a power law and quantified the excess
attenuation relative to this continuum as \citep[see][]{Shivaei2022,Witstok2023}:
\begin{equation}
A_{\lambda,\rm bump} = -2.5\log_{10}\left(\frac{F_\lambda}{F_{\lambda,\rm cont}}\right).
\end{equation}
We then fitted the resulting attenuation excess over the rest-frame $1950\mbox{--}2580\,\AA$ interval encompassing the feature with a Drude profile, i.e., the functional form commonly used to describe the broad $2175\,\AA$ absorption feature. The best-fitting profile provides a good description of the observed depression, with no significant coherent structure remaining in the residuals (Fig.\,\ref{fig:drude}).

To quantify the statistical significance of the feature,
we followed an approach analogous to that adopted by
\citet{Witstok2023} and \citet{Ormerod2025}. We applied a running-median
filter to the spectrum and estimated the corresponding uncertainty through
Monte Carlo perturbations of the native spectral pixels according to their
formal errors. We then measured the integrated negative flux excess of the
smoothed spectrum relative to the locally fitted power-law continuum over the
wavelength interval used to characterize the bump. This yields a formal
negative-excess significance of $17.6\sigma$.\footnote{We quantify the significance of the feature from the integrated normalized flux deficit relative to the no-bump power-law continuum. Defining
\begin{equation}
R_i = \frac{F_{\lambda,i}}{F_{\lambda,{\rm PL},i}},
\end{equation}
we compute
\begin{equation}
S_{\rm bump} =
\frac{\sum_i (1-R_i)}
{\sqrt{\sum_i \sigma_{R,i}^{2}}},
\end{equation}
where $\sigma_{R,i}$ is derived from bootstrap realizations of the
running-median spectrum. This yields a formal significance of
$S_{\rm bump}=17.6\sigma$.} Independently, we compared the
goodness of fit obtained with the power-law continuum alone and with the
addition of a Drude component over the same local interval. The fit improves
from $\chi^2_{\rm PL}=474$ to $\chi^2_{\rm PL+Drude}=8.2$, providing strong
statistical support for the presence of the broad attenuation feature
(Fig.\,\ref{fig:drude}).

The local Drude fit yields a peak attenuation
$A_{\rm max}=0.21\pm0.02\,\mathrm{mag}$, a characteristic width
${\rm FWHM}=413\pm54\,\AA$, and a central wavelength
$\lambda_{\rm max}=2132\pm17\,\AA$. 

The inferred central wavelength is blueshifted relative to the
canonical $2175\,\AA$ feature observed in the Milky Way, with the
offset corresponding to a significance of $\simeq2.5\sigma$ based
on the uncertainty of the local Drude fit. The measured width is
instead consistent with the broad distribution of FWHM values seen
along Galactic sightlines. This purely spectral analysis
therefore identifies the displaced central wavelength as the most distinctive aspect of the bump.

The adopted fitting interval excludes the \ciii$\lambda\lambda1907,1909$ emission. To assess whether this feature
could nevertheless affect the inferred bump position, we repeated the
fit extending the wavelength interval down to $1850\,\AA$. Explicitly
masking the \ciii{} region ($1875\mbox{--}1940\,\AA$) yields
$\lambda_{\rm max}=2136\pm15\,\AA$, while including this region yields
$\lambda_{\rm max}=2142\pm13\,\AA$. Both values are consistent with the
fiducial measurement, indicating that \ciii{} emission has only
a minor effect on the inferred bump position.

\begin{figure}
\includegraphics[width=\columnwidth]{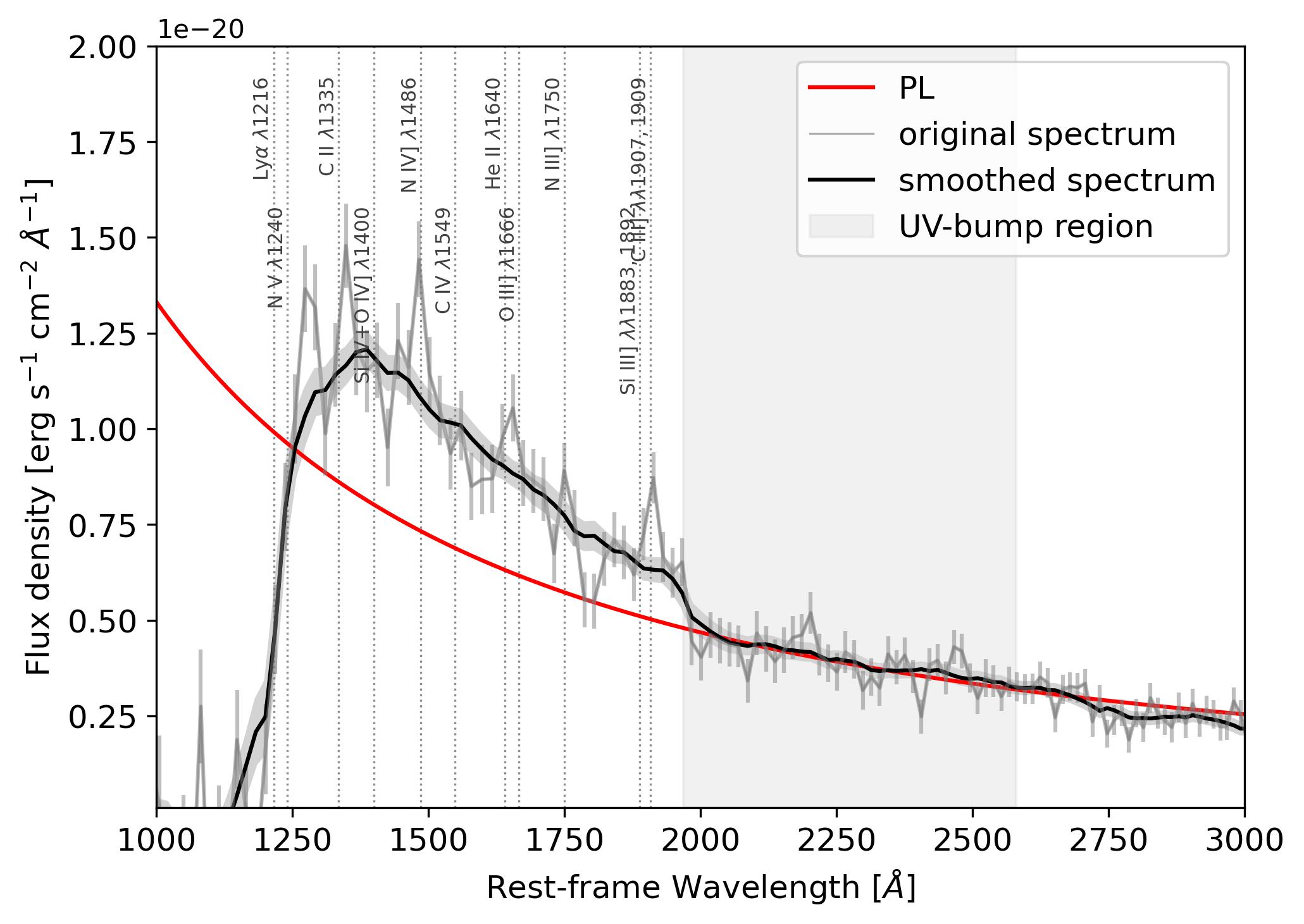}
\caption{
Rest-frame UV spectrum used to assess the broad continuum depression
associated with the UV bump. The original NIRSpec/PRISM spectrum is
shown in gray and the smoothed spectrum in black. The red curve shows
the power-law continuum fitted at $\lambda_{\rm rest}>2000\,\AA$ and
extrapolated toward shorter wavelengths. The gray shaded region marks
the wavelength interval $1968$--$2580\,\AA$ adopted to characterize
the UV bump \citep[as in][]{Witstok2023}. Vertical dotted lines indicate the expected wavelengths
of prominent rest-frame UV spectral features, including Ly$\alpha$,
N\,V, C\,II, Si\,IV+O\,IV], N\,IV], C\,IV, He\,II,
O\,III], N\,III], Si\,III], and C\,III].
Although individual spectral features may contribute locally to the
UV spectrum, they cannot reproduce the broad spectral curvature and
continuum recovery observed blueward of the UV-bump region.
}
    \label{fig:2000}
\end{figure}

As a complementary robustness test, we investigated whether the observed
continuum blueward of the absorption feature could instead be produced by
unresolved or blended emission lines. We fitted a power-law continuum using
only the spectrum at $\lambda_{\rm rest}>2000\,\AA$, thereby excluding the
blue UV continuum from the fit, and extrapolated the resulting continuum
toward shorter wavelengths (Fig.\,\ref{fig:2000}). The extrapolated power law
does not reproduce the pronounced rise of the observed spectrum blueward of
$\sim1900\,\AA$. Figure~\ref{fig:2000} also marks the expected wavelengths of
the main rest-frame UV spectral features that could contribute in this region.
While individual features may locally affect the spectrum, the observed excess
extends over a broad wavelength interval and is not confined to the expected
positions of these transitions. Unresolved line emission alone is therefore
unlikely to account for the observed continuum shape. This test is independent
of the Drude-profile characterization above and supports the interpretation of
the feature as broad continuum attenuation rather than line contamination.

\subsection{Emission-line measurements}
\label{sec:emission_lines}
The NIRSpec spectra reveal several rest-frame optical emission
lines, including \oii\,$\lambda\lambda3726,3729$,
\neiii\,$\lambda3869$, H$\gamma$, \oiii\,$\lambda4363$,
H$\beta$, and the \oiii\,$\lambda\lambda4959,5007$ doublet
(Fig. \ref{fig:spectrum}). The medium-resolution G395M spectrum, shown in
Fig. \ref{fig:spectrum_grism}, provides a higher resolution view of the \oii,
\neiii, H$\gamma$, and \oiii\,$\lambda4363$ emission
features.

We measured the emission-line fluxes by first estimating the underlying
continuum locally around each spectral feature and subtracting it from
the observed spectrum. The continuum-subtracted emission lines were
then fitted with Gaussian profiles. Line fluxes were obtained by
integrating the best-fitting Gaussian models, with the corresponding
uncertainties propagated from the spectral uncertainties. We considered
as significant detections lines with an integrated signal-to-noise
ratio ${\rm S/N}>3$.

Whenever covered by both NIRSpec configurations, we performed the
measurements independently on the PRISM and G395M spectra. This provides
an important consistency check between the low- and medium-resolution
observations. In particular, \oii\,$\lambda\lambda3726,3729$,
\neiii\,$\lambda3869$, and H$\gamma$ are independently detected in
both spectra, with fluxes consistent within the uncertainties.

The medium-resolution G395M spectrum also provides a tentative
detection of the auroral \oiii\,$\lambda4363$ line at approximately
$3\,\sigma$. If confirmed, this feature provides the possibility of
constraining the electron temperature and hence the gas-phase
metallicity through the direct $T_{\rm e}$ method. We therefore regard
the \oiii\,$\lambda4363$ measurement as tentative in the following
analysis.

At longer wavelengths, the PRISM spectrum shows strong H$\beta$ and
\oiii\,$\lambda\lambda4959,5007$ emission, detected at high
significance. The measured emission-line fluxes and their uncertainties
are summarized in Table~\ref{tab:lines}.

\begin{table}
\centering
\caption{Emission-line measurements for the Black Pearl. Fluxes are given in units of $10^{-21}\,\mathrm{erg}\,\mathrm{s}^{-1}\,\mathrm{cm}^{-2}$. Only lines detected at integrated $\mathrm{S/N}>3$ are reported as detections.}
\label{tab:lines}
\begin{tabular}{lccc}
\hline\hline
Line & Instrument & Flux & S/N \\
\hline
\oii\,$\lambda\lambda$3726,3729 & PRISM & $532.4\pm158.7$ & 3.4 \\
                                       & G395M & $761.6\pm117.1$ & 6.5 \\

\neiii\,$\lambda3869$            & PRISM & $1066.5\pm224.7$ & 4.7 \\
                                    & G395M & $622.4\pm200.0$ & 3.1 \\

H$\gamma$\,$\lambda4340$            & PRISM & $622.2\pm194.9$ & 3.2 \\
                                    & G395M & $700.0\pm156.9$ & 4.5 \\

\oiii\,$\lambda4363$             & G395M & $476.5\pm156.9$ & 3.0 \\

H$\beta$ $\lambda4861$             & PRISM & $1630.7\pm150.0$ & 10.9 \\

\oiii\,$\lambda4959$             & PRISM & $3854.8\pm196.0$ & 19.7 \\

\oiii\,$\lambda5007$             & PRISM & $10895.3\pm196.0$ & 55.6 \\
\hline
\end{tabular}
\end{table}

\begin{figure*}
\includegraphics[width=18cm]{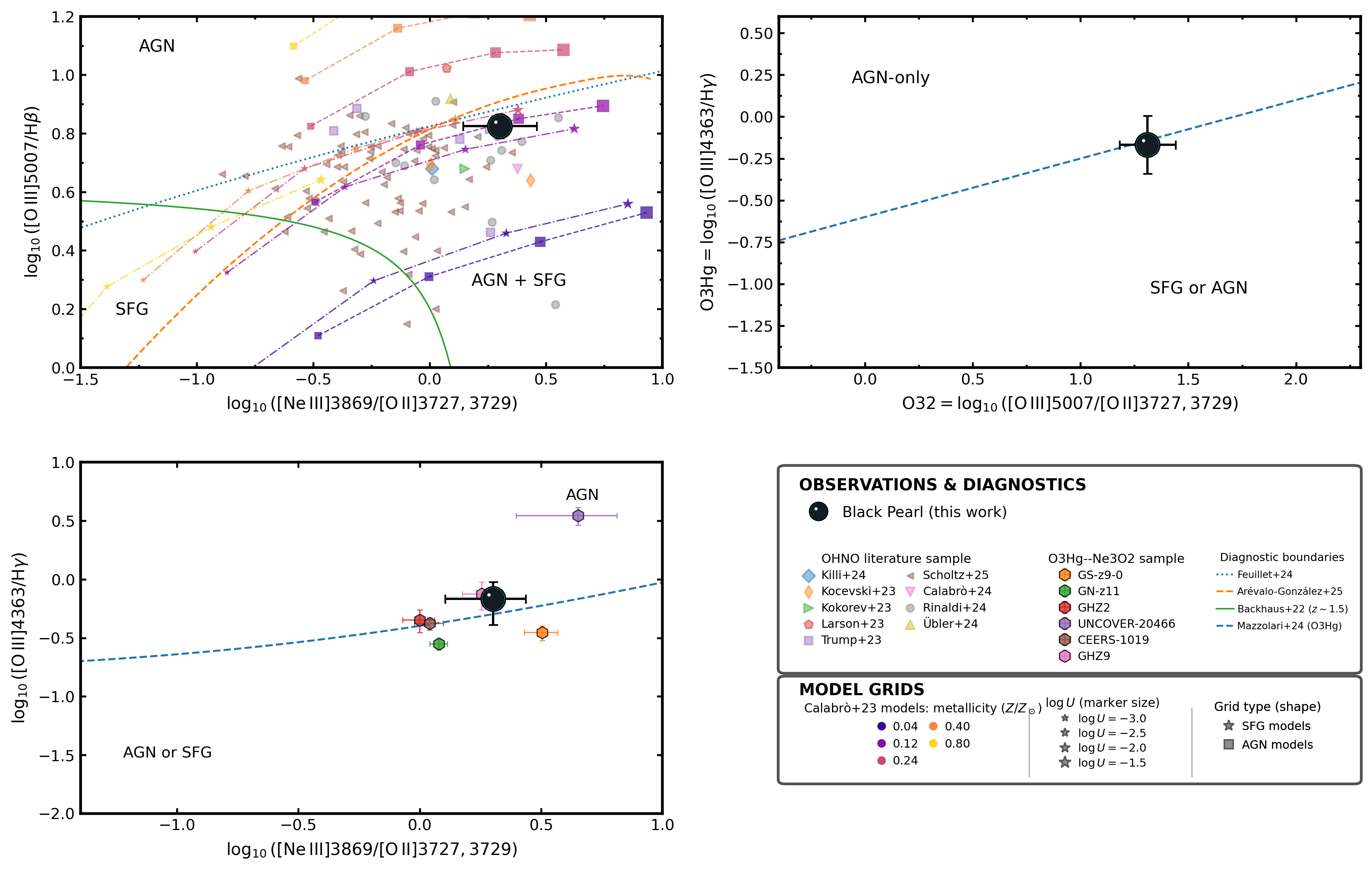}
\caption{Rest-frame optical emission-line diagnostic diagrams for the
Black Pearl (black circular symbol).
\textit{Top left:} OHNO diagram, compared with literature measurements,
AGN/star-forming galaxies (SFGs) demarcation curves from \citet{Feuillet2024},
\citet{ArevaloGonzalez2025}, and \citet{Backhaus2022}, and the SFG and AGN
photoionization grids of \citet{Calabro2023}. The colored symbols show
recent literature measurements at $z\gtrsim4$ from \citet{Killi2024},
\citet{Kocevski2023}, \citet{Kokorev2023}, \citet{Larson2023},
\citet{Trump2023}, \citet{Uber2024}, \citet{Scholtz2025},
\citet{Calabro2024}, and \citet{Rinaldi2024}.
\textit{Top right:} O3Hg--O32 diagnostic.
\textit{Bottom left:} O3Hg--Ne3O2 diagnostic. The dashed blue curves in the
latter two panels show the AGN/SFG boundaries from \citet{Mazzolari2024}.
The comparison measurements shown in the bottom-left panel are from
GS-z9-0 \citep{Curti2025}, GN-z11 \citep{Bunker2024},
GHZ2 \citep{Castellano2024}, UNCOVER-20466 \citep{Kokorev2023},
CEERS-1019 \citep{Larson2023}, and GHZ9 \citep{Napolitano2025}.
The two boxed legends on the right identify the observational samples
and diagnostic boundaries (upper box) and the photoionization model
grids (lower box). In the latter, marker colors encode metallicity,
marker sizes indicate the ionization parameter $\log U$, and marker
shapes distinguish SFG and AGN models.}
    \label{fig:diagnostics}
\end{figure*}

\subsection{Ionization properties and spectral classification}\label{sec:agn}

To investigate the nature of the ionizing source, we considered three
complementary rest-frame optical emission-line diagnostics, shown in
Fig.\,\ref{fig:diagnostics}. The top-left panel shows the OHNO diagram
\citep{Backhaus2022}, based on \oiii\,$\lambda5007$/H$\beta$ and
\neiii\,$\lambda3869$/\oii\,$\lambda\lambda3726,3729$. The top-right
and bottom-left panels show the two diagnostics proposed by
\citet{Mazzolari2024}, which exploit the temperature-sensitive auroral
\oiii\,$\lambda4363$ line: O3Hg--O32 and O3Hg--Ne3O2, respectively.

We caution that the empirical AGN/SF demarcations adopted in these diagrams were largely calibrated at substantially lower redshifts and may not remain applicable at $z=9.3335$. In particular, the evolution of nebular excitation and ionizing spectra can shift the locus of star-forming galaxies toward regions conventionally associated with AGN activity \citep[][]{Backhaus2025,Rinaldi2025}.
We therefore use these diagnostics to characterize the excitation conditions and assess the plausibility of an AGN contribution, rather than to establish a definitive spectral classification.

In the OHNO diagram (top-left panel), the source lies at high excitation, in a region where the different AGN/SF demarcations do not provide a unique classification.
We note that recent AGN photoionization models further highlight the limited
discriminatory power of the \oiii/H$\beta$ versus
\neiii/\oii{} diagram, particularly under the low-metallicity
conditions relevant to early galaxies \citep{Plat2026}.
While some empirical
criteria extend the AGN locus toward this part of the diagram,
photoionization models predict a substantial overlap between AGN
and extreme star-forming populations. Indeed, compact, metal-poor
starbursts characterized by high ionization parameters or
density-bounded H\,II regions can reach similarly high excitation
ratios \citep[e.g.,][]{Larson2023,Tripodi2024,Scholtz2025}.

This ambiguity is also apparent when comparing the source with the
photoionization grids of \citet{Calabro2023} shown in Fig. \ref{fig:diagnostics}.
The grids span different metallicities and ionization parameters
for both star-forming and AGN models, illustrating the strong
degeneracy between the two ionizing scenarios in this region of
the OHNO plane. 
In particular, the ionization parameter and metallicity independently inferred from the SED modeling, $\log U=-1.45^{+0.30}_{-0.45}$ and
$Z/Z_\odot=0.20^{+0.63}_{-0.13}$ (Sect.\,\ref{sed_fitting_results}; Table \ref{tab:posterior}), favor the high-ionization, subsolar metallicity region of the model grids. Within this range of physical parameters, both star-forming and AGN photoionization models can reproduce line ratios comparable to those observed.
The position of the source in the OHNO diagram therefore indicates a hard and highly ionized radiation field, but does not by itself require an AGN-dominated spectrum (see also the discussion in \citealt{Iani2026}).

The diagnostics involving the auroral \oiii\,$\lambda4363$ line provide
an independent, although less constraining, test. In the O3Hg--O32
(top-right panel) and O3Hg--Ne3O2 (bottom-left panel) diagrams, the
galaxy lies close to the empirical/model demarcation between AGN and
star-forming systems. Given the $\sim 3\,\sigma$ significance of the
\oiii\,$\lambda4363$ detection, the corresponding uncertainties are
substantial and prevent an unambiguous classification.

Taken together, the available line-ratio diagnostics are
consistent with intense star formation and allow for a possible
AGN contribution, but do not provide an unambiguous separation
between the two ionizing scenarios.
Importantly, however, the source is inconsistent with the highly
ionized regime occupied by the most extreme AGN-dominated systems.
We further explore the nature of the ionizing source and the nebular geometry using depth-dependent photoionization models (Appendix~\ref{app:ionization}; Fig.\,\ref{fig:borja}). The observed strong-line ratios are consistent with stellar photoionization in a moderately matter-bounded configuration, potentially allowing a modest escape of ionizing photons. This interpretation is also compatible with the escape fraction of a few percent inferred indirectly from the UV continuum slope. While an AGN contribution cannot be excluded, the additional modeling does not require an AGN-dominated ionizing spectrum.

We discuss the implications for the interpretation of the UV continuum and the origin of the broad absorption feature in Sect.~\ref{sec:discussion}.


\section{SED fitting}\label{sed_fitting}
\subsection{SED-fitting methodology}
To obtain further independent evidence for the presence of the UV bump, we adopt an alternative Bayesian approach that is complementary to the method presented in Section~\ref{identification}. We performed SED fitting with the Bayesian Analysis of Galaxies for Physical Inference and Parameter EStimation code \citep[\texttt{BAGPIPES},][]{Carnall2018}, which simultaneously fits spectroscopic and photometric observations within a Bayesian framework. The parameter space was explored using the \texttt{MultiNest} nested sampling algorithm \citep{Feroz2008}. We adopted the \citet{Bruzual2003} stellar population synthesis models together with a \citet{Kroupa2002} IMF. The fitting procedure compares the model predictions with the observed data and recovers the posterior probability distribution of the model parameters. Throughout this work, the inferred parameters are quoted as the posterior median together with their $68\%$ credible intervals, corresponding to the 16th, 50th, and 84th percentiles of the posterior distribution. As in \citet{Markov2023}, we modified the \texttt{BAGPIPES} source code to implement a parameterization of the attenuation curve. This approach is particularly well suited to high-redshift galaxies, where the dust attenuation properties remain largely unconstrained and may differ significantly from those observed in the local Universe. Consequently, adopting conventional attenuation laws calibrated on nearby galaxies may introduce systematic biases in the inferred physical properties of distant systems \citep{Maiolino2004,Li2008,Markov2023,Markov2025}. A flexible attenuation law therefore provides a more appropriate framework for investigating the dust properties of galaxies in the early Universe. As discussed by \citet{Markov2023}, the main drawback of this approach is the introduction of additional free parameters compared to conventional attenuation prescriptions, substantially increasing the dimensionality of the parameter space. Robustly constraining these parameters therefore requires, in the ideal case, both high-quality spectroscopy and broadband photometry to reduce parameter degeneracies and fully exploit the Bayesian inference. The principal advantage, however, is that the attenuation curve is not restricted to any predefined functional form. This flexibility enables us to place further constraints on  the features of the UV bump in the galaxy at this unprecedented redshift.

Since our primary goal is to further characterize the presence and strength of the UV bump in this galaxy, we adopt the analytical attenuation prescription proposed by \citet{Li2008} and modified versions. Owing to its flexibility, this formulation can reproduce attenuation curves similar to several well-established extinction and attenuation laws, including those of \citet{Cardelli1989} for the Milky Way (MW), \citet{Gordon2003} for the Small Magellanic Cloud (SMC), \citet{Weingartner2001} for the Large Magellanic Cloud (LMC), and \citet{Calzetti2000}, while also allowing for attenuation curves with nonstandard shapes.

\citet{Ormerod2025} introduced a modification to the original prescription
that allows the characteristic width of the UV bump to differ from the
standard Galactic value; see their Equation (2). In their implementation, both
the central wavelength and the FWHM of the UV bump are fixed to $\lambda_{\rm max}=2175\,\AA$ and $\mathrm{FWHM}=250\,\AA$, respectively. We further extend this
prescription by treating both $\lambda_{\rm max}$ and FWHM as free
parameters in the fit:

\begin{equation}
\begin{aligned}
\frac{A_{\lambda}}{A_V}
=&\;
\frac{c_1}
{\left(\lambda/0.08\right)^{c_2}
+\left(0.08/\lambda\right)^{c_2}
+c_3} \\[0.2cm]
&+
\frac{233\left[1-\dfrac{c_1}
{6.88^{c_2}+0.145^{c_2}+c_3}
-\dfrac{c_4}{x}\right]}
{\left(\lambda/0.046\right)^2
+\left(0.046/\lambda\right)^2
+90} \\[0.2cm]
&+
\frac{c_4}
{\left(\lambda/\lambda_{\rm max}\right)^2
+\left(\lambda_{\rm max}/\lambda\right)^2
+\left(\text{FWHM}/\lambda_{\rm max}\right)^2
-2},
\end{aligned}
\label{eq:drude}
\end{equation}

where $\lambda$ is expressed in $\micron$ and $x$ is given by:

\begin{equation}
x =
\left(\frac{0.55}{\lambda_{\rm max}}\right)^2
+
\left(\frac{\lambda_{\rm max}}{0.55}\right)^2
+
\left(\frac{\text{FWHM}}{\lambda_{\rm max}}\right)^2
-2.
\label{eq:drude_norm}
\end{equation}
We fit simultaneously $c_1$, $c_2$, $c_3$, $c_4$, $\lambda_{\rm max}$, and FWHM.  The amplitude of the UV bump is controlled by the parameter $c_4$, with larger values corresponding to stronger absorption features. For reference, the MW and LMC extinction curves are characterized by $c_4=0.0519$ and $c_4=0.0221$, respectively, in the original parameterization of \cite{Li2008}. 
This additional flexibility enables us to independently constrain the position and width of the UV bump from the observed galaxy spectra and photometry, rather than imposing the canonical MW values. Using the attenuation prescriptions described above, we find evidence for a UV bump feature. Here, we report the results from the more flexible parameterization in Eqn. \eqref{eq:drude}, motivated by the evidence for a blueshift of the bump central wavelength found in our previous test (Section~\ref{identification}).







To assess the dependence of our results on the assumed star-formation histories (SFH), we considered two different SFH parameterizations. The first is a delayed exponentially declining SFH, which has been widely adopted in studies of high-redshift galaxies \citep[e.g.,][]{Furtak2023}. The second is the nonparametric continuity SFH, in which the SFH is described by a series of piecewise-constant star formation rates (SFRs), with continuity priors enforcing smooth transitions between adjacent time bins. This parameterization provides greater flexibility to reproduce the often complex star formation histories of galaxies \citep{Leja2019,Tacchella2022}. We adopted a three-bin nonparametric SFH to limit the number of free parameters and reduce potential parameter degeneracies.

\begin{figure}
\centering
\includegraphics[width=\columnwidth]{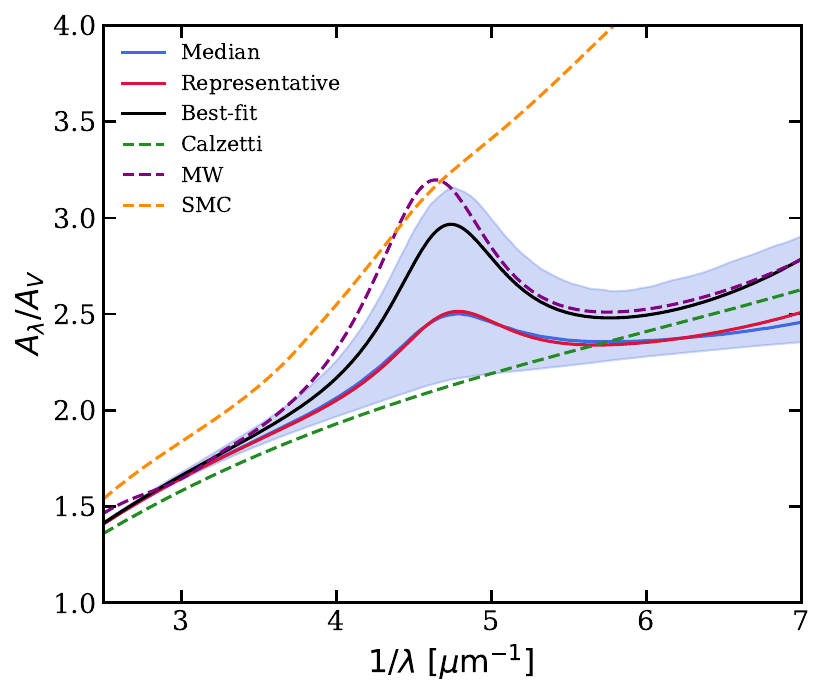}
\caption{Posterior attenuation curves obtained with the nonparametric SFH model. The blue curve shows the median attenuation curve, while the shaded blue region encloses the 16th--84th percentiles at each wavelength. The red curve corresponds to the representative posterior model, while the black curve shows the maximum-likelihood (best-fitting) posterior model (see Section \ref{sed_fitting_results} for details). For comparison, standard attenuation and extinction curves are also shown.}
\label{fig:dust_curves}
\end{figure}

Table~\ref{tab:priors} summarizes the free parameters and priors adopted in the \texttt{BAGPIPES} fitting. The redshift is kept fixed at $z_{\rm spec}=9.3335$. Unless otherwise stated, all priors correspond to uniform distributions within the quoted intervals. The stellar population synthesis models, IMF, nebular emission, intergalactic medium attenuation \citep{Inoue2014}, and Drude dust attenuation parameterization were kept identical for both SFH prescriptions. We obtain consistent results with the two SFH parameterizations, both in terms of the inferred galaxy properties and the properties of the UV bump. The attenuation curve and the corresponding results obtained with the nonparametric SFH are presented in Fig.\,\ref{fig:dust_curves} and Table\,\ref{tab:posterior}.

\begin{table}
\caption{Free parameters and priors adopted in the \texttt{BAGPIPES} SED fitting. Uniform priors were assumed unless otherwise stated.}
\label{tab:priors}
\centering
\begin{tabular}{lc}
\hline\hline
Parameter & Prior / Value \\
\hline

\multicolumn{2}{l}{\textbf{Dust attenuation}}\\
\hline
$A_V$ & $[0,\;6]$ \\
$c_1$ & $[0,\;75]$ \\
$c_2$ & $[2,\;20]$ \\
$c_3$ & $[-2,\;75]$ \\
$c_4$ & $[-0.005,\;0.1]$ \\
$\lambda_{\rm max}$ ($\micron$) & $[0.2050,\;0.2300]$ \\
$\text{FWHM}$ ($\micron$) & $[0.0200,\;0.0500]$ \\
\hline

\multicolumn{2}{l}{\textbf{Delayed SFH}}\\
\hline
Age (Gyr) & $[0.001,\;0.55]$ \\
$\tau$ (Gyr) & $[0.01,\;3]$ \\
$\log(M_{\rm formed}/M_\odot)$ & $[6.5,\;10.5]$ \\
$Z/Z_\odot$ & $[0,\;2.5]$ \\
\hline

\multicolumn{2}{l}{\textbf{nonparametric continuity SFH}}\\
\hline
Bin edges (Myr) & $[0,\;10,\;50,\;t_{\rm U}-10]$ \\
$\Delta\log{\rm SFR}_1$ & $[-10,\;10]$ \\
$\Delta\log{\rm SFR}_2$ & $[-10,\;10]$ \\
\hline

\multicolumn{2}{l}{\textbf{Nebular emission}}\\
\hline
$\log U$ & $[-4,\;-1]$ \\
\hline
\end{tabular}
\end{table}

\subsection{SED-fitting results}\label{sed_fitting_results}
Figure \ref{fig:dust_curves} shows three different representations of the posterior attenuation curve from \texttt{BAGPIPES}. The blue curve corresponds to the median attenuation curve, obtained by computing the median value of $A_\lambda/A_V$ independently at each wavelength from the posterior distribution of attenuation curves, along with the 68\% credible interval at each wavelength (blue band) and that describes the central tendency of the posterior distribution and its associated uncertainty.
The red curve shows a representative posterior model. This curve is obtained by selecting the posterior sample whose parameter vector has the smallest normalized distance to the vector of posterior median values. Consequently, the entire attenuation curve is generated from a single posterior realization, ensuring that all fitted parameters (Table \ref{tab:priors}) correspond to one physically self-consistent model that is representative of the posterior distribution.
Finally, the black curve corresponds to the best-fitting model, defined as the posterior sample with the highest likelihood, providing the closest match to the observed data. Each curve highlights a different statistical property of the inferred posterior distribution while consistently supporting the presence of the UV bump.

\begin{figure*}[h]
\centering
\includegraphics[width=18cm]{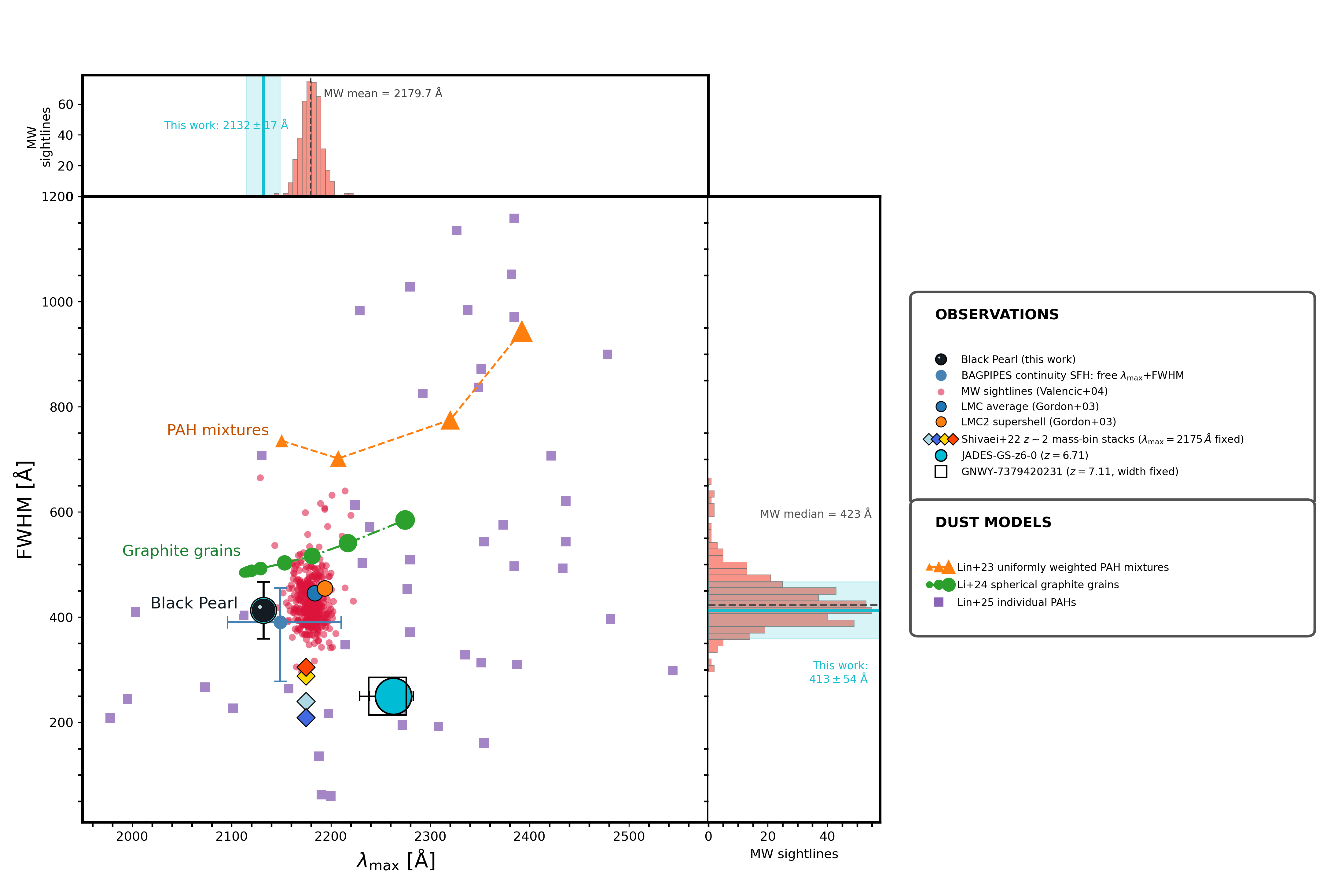}
\caption{UV-bump profile space. The bump central wavelength,
$\lambda_{\max}$, is plotted against its full width at half maximum
(FWHM). Red points represent individual Milky Way (MW) sightlines
from \citet{Valencic2004}. The marginal histograms show their
$\lambda_{\max}$ distribution (top) and FWHM distribution (right);
the dashed reference lines indicate the MW mean
$\lambda_{\max}=2179.7\,\AA$ and median FWHM $=423\,\AA$.
The cyan vertical and horizontal bands indicate the $1\,\sigma$
intervals of the direct spectroscopic measurement for the Black Pearl,
$\lambda_{\max}=2132 \pm 17\,\AA$ and
FWHM $=413 \pm 54\,\AA$, respectively.
The Black Pearl at $z_{\rm spec}=9.3335$ is marked by the large black circle,
with its $1\,\sigma$ error bars; the blue point and error bars show
the joint spectrophotometric \texttt{BAGPIPES}
constraints obtained with the nonparametric continuity
star-formation history and with both bump parameters free.
The blue filled circle and orange filled circle indicate the
average LMC and LMC2 supershell
measurements from \citet{Gordon2003}, respectively.
Four colored diamonds at $\lambda_{\max}=2175\,\AA$
represent the $z\sim2$ stellar-mass-bin stacks from
\citet{Shivaei2022}; the central wavelength was fixed in those fits.
The cyan filled circle marks JADES-GS-z6-0 ($z=6.71$), while
the open square marks GNWY-7379420231 ($z=7.11$), for which
the bump width was fixed.
Marker sizes for the observational measurements are proportional to the peak bump attenuation, $A_{\max}$.
The observational datasets and theoretical models are identified in the two legend boxes on the right.
The theoretical predictions include orange triangles connected
by a dashed line for uniformly weighted PAH mixtures from
\citet{Lin2023}, green circles connected by a dash-dotted line
for spherical graphite grains from \citet{Li2024}, and purple
squares for individual PAHs from \citet{Lin2025}.
Increasing marker sizes along the orange and green model sequences
indicate increasing PAH-mixture size and graphite-grain radius,
respectively. The PAH-mixture and graphite-grain sequences,
together with the Black Pearl, are labeled directly in the main panel.}
\label{fig:bump_profile_space}
\end{figure*}

Despite probing different statistical properties of the posterior,
all three representations consistently recover a UV bump (Fig.\,\ref{fig:dust_curves}).
Importantly, although the adopted attenuation prescription explicitly allows for a UV bump, its amplitude, central wavelength, and width are free parameters constrained by the joint spectrophotometric fit.
Moreover, a similar feature is recovered independently when adopting
either the delayed or the nonparametric SFH. For the delayed-$\tau$ SFH, the resulting posterior constraints are
$\lambda_{\rm max}=2147^{+51}_{-59}\,\AA$ and
${\rm FWHM}=392^{+109}_{-61}\,\AA$, while the nonparametric continuity
SFH yields $\lambda_{\rm max}=2168^{+77}_{-73}\,\AA$ and
${\rm FWHM}=378^{+78}_{-99}\,\AA$. The two  SFH
parameterizations therefore recover remarkably similar bump profiles, providing an independent line of evidence supporting the spectral analysis
of Sect.\,\ref{identification}, where the continuum depression was identified and
characterized directly from the NIRSpec spectrum.

As shown in Fig.\,\ref{fig:bump_profile_space}, these constraints are
fully consistent with those obtained from the independent local analysis
of the NIRSpec spectrum,
$\lambda_{\rm max}=2132\pm17\,\AA$ and
${\rm FWHM}=413\pm54\,\AA$. Thus, despite the substantially larger
parameter space and the simultaneous modeling of the stellar populations,
dust attenuation, nebular emission, and broadband photometry, BAGPIPES
independently recovers both the central wavelength and width inferred
directly from the spectrum. 
This agreement provides an additional robustness check on the identification of the feature and shows that the bump parameters inferred from the full spectrophotometric modeling are consistent with those obtained from the direct spectral analysis.
For completeness, Appendix \ref{sed-nobump-sibump} (Fig. \ref{fig:fitfit}) compares the best-fitting \texttt{BAGPIPES} spectra obtained with and without the Drude UV-bump component, adopting the reference nonparametric continuity SFH and the flexible attenuation prescription described above. This comparison illustrates the effect of including the bump on the modeled UV continuum.

Table \ref{tab:posterior} summarizes the posterior constraints obtained from the SED fitting using the nonparametric continuity SFH.  The parameter estimates obtained with the delayed SFH are fully consistent within the quoted uncertainties and therefore are not listed separately. 
The inferred stellar mass is
$\log(M_\star/M_\odot)=8.35^{+0.09}_{-0.08}$.
The fit also requires nonnegligible dust attenuation,
$A_V=0.335^{+0.123}_{-0.118}$ mag, while the inferred ionization parameter, $\log U=-1.432^{+0.281}_{-0.401}$,  favors highly ionized conditions,
although, as discussed below, its posterior approaches the upper
boundary of the adopted prior. The flexible dust parameterization allows us to directly constrain the shape of the attenuation curve and, in particular, the strength of the UV bump.

\begin{table}
\caption{Posterior constraints obtained from the SED fitting using the nonparametric continuity SFH. Quoted values correspond to the posterior median together with the 68\% credible interval (16th--84th percentiles).}
\label{tab:posterior}
\centering
\begin{tabular}{lc}
\hline\hline\noalign{\vskip 1pt}
Parameter & Posterior value  \\
\hline\noalign{\vskip 1pt}
$A_V$ & $0.335^{+0.123}_{-0.118}$  \\
$c_1$ & $40.27^{+21.15}_{-22.49}$  \\
$c_2$ & $9.47^{+5.37}_{-3.70}$  \\
$c_3$ & $35.00^{+24.31}_{-24.32}$  \\
$c_4$ & $0.0121^{+0.0212}_{-0.0116}$  \\
$\lambda_{\rm max}$ ($\micron$) & $0.2168^{+0.0077}_{-0.0073}$ \\
$\mathrm{FWHM}$ ($\micron$) & $0.0378^{+0.0078}_{-0.0099}$  \\
\hline\noalign{\vskip 1pt}
$\log_{10}(M_\star/M_\odot)$ & $8.35^{+0.09}_{-0.08}$  \\
SFR ($M_\odot\,{\rm yr}^{-1}$) & $2.364^{+0.540}_{-0.424}$ \\
\hline\noalign{\vskip 1pt}
Metallicity ($Z/Z_\odot$) & $0.247^{+0.638}_{-0.164}$  \\
\hline\noalign{\vskip 1pt}
$\log_{10} U$ & $-1.432^{+0.281}_{-0.401}$ \\
\hline\noalign{\vskip 1pt}
$\Delta\log{\rm SFR}_1$ & $3.547^{+3.988}_{-4.553}$  \\
$\Delta\log{\rm SFR}_2$ & $6.535^{+2.200}_{-2.681}$ \\
\hline
\end{tabular}
\end{table}

We caution that some parameters of the flexible SED model are only
weakly constrained. In particular, the posterior of $c_4$ extends
toward very small values, while the ionization parameter approaches
the upper boundary of the adopted prior, indicating that $\log U$
is prior-limited toward high ionization. The
$\Delta\log({\rm SFR})$ parameters of the nonparametric SFH are
also poorly constrained. We therefore do not attach strong physical
significance to these individual parameters.

Overall, the SED modeling therefore indicates a relatively low-mass,
actively star-forming galaxy with no-negligible dust attenuation.
Most importantly for the present work, the inferred attenuation curve
independently supports the presence of the broad UV feature identified
directly in the NIRSpec spectrum, with a central wavelength and width
consistent with the independent local spectral analysis. We discuss
the implications of its shape and central wavelength in Sect.\,\ref{sec:discussion}.

\section{Discussion}
\label{sec:discussion}

\subsection{The nature of the UV bump at $z=9.3335$}
The evidence for a UV attenuation bump in the Black Pearl is supported by two
complementary analyses. First, the feature is identified directly in the NIRSpec spectrum. 
The local continuum analysis presented in Sect.~\ref{identification} shows that the broad depression
is detected at a formal $17.6\sigma$ negative-flux-excess significance and is very strongly preferred over a featureless power-law continuum. The feature is well described by a local Drude profile, with no significant coherent structure remaining in the residuals. 
This purely spectroscopic approach yields $\lambda_{\rm max}=2132\pm17\,\AA$, ${\rm FWHM}=413\pm54\,\AA$, and a peak attenuation of $A_{\rm max}=0.21\pm0.02$ mag.

Second, the joint spectrophotometric modeling independently favors an attenuation curve containing a UV bump.
Importantly, the flexible Drude attenuation prescription adopted in the \texttt{BAGPIPES} analysis does not impose a Milky-Way-like extinction curve, nor does it require the presence of a bump.
Instead, its strength is allowed to vary as part of the Bayesian inference. 
As shown in Fig.\,\ref{fig:dust_curves}, the median posterior attenuation curve exhibits a UV bump around the expected wavelength, although the $1\sigma$  region encompasses a broad range of
bump strengths, including curves with a very weak or negligible feature. The bump is also present in a representative posterior realization and in the maximum-likelihood solution.
 Moreover, it is recovered independently when adopting either the delayed or
the nonparametric SFH. 
The SED modeling therefore provides an independent, although less stringent, consistency check of the spectroscopically identified feature, without requiring a particular star-formation history or imposing a predefined local attenuation law.

The consistency between these two approaches is particularly important. The local spectral analysis establishes that the observed continuum depression has the characteristic broad
profile expected for a UV attenuation feature, whereas the full spectrophotometric modeling yields attenuation curves consistent with the spectroscopically identified bump,
although its strength remains poorly constrained.
More importantly, when both the central wavelength and the width of the bump are allowed to vary freely in the \texttt{BAGPIPES} fit, the resulting posterior distribution in the $\lambda_{\rm max}$--FWHM plane overlaps the constraints obtained from the independent local fit to the NIRSpec spectrum (Fig.\,\ref{fig:bump_profile_space}). 
Thus, the two analyzes not only independently support the presence of a UV bump but also recover consistent constraints on its profile despite their very different modeling approaches. 
In particular, the \texttt{BAGPIPES} posterior has a median central wavelength blueward of the canonical Milky Way value and is fully consistent with the more precise measurement obtained from the direct spectral analysis. Although the broader \texttt{BAGPIPES} constraint alone does not require a blueshift relative to the Milky Way value, this agreement indicates that the displaced central wavelength inferred from the local fit is compatible with the full spectrophotometric modeling.

Beyond establishing the presence of the bump, its detailed shape provides a means of probing the properties of the grain
population at Cosmic Dawn. The clearest difference with respect to the canonical Milky Way bump is its central wavelength:
the feature is displaced toward shorter wavelengths, with $\lambda_{\rm max}=2132\pm17\,\AA$ compared with the
Galactic mean near 2179.7~\AA. In contrast, its measured width, ${\rm FWHM}=413\pm54\,\AA$, is fully consistent with the broad
distribution of widths observed along the Milky Way sightlines and with the Galactic median of $\simeq423\,\AA$ (Fig.\,\ref{fig:bump_profile_space}).

However, we caution that the detailed shape of an attenuation curve does not provide a unique mapping to the properties
of the underlying grain population. Radiative-transfer effects, optical depth, and the relative geometry of stars and dust can
modify the observed attenuation curve and introduce degeneracies between different dust compositions \citep[e.g.,][]{Witt1992,Sommovigo2026}. At the same time, UV-bump-related properties retain sensitivity to the underlying
grain composition, making the displaced profile observed here a potentially valuable, although not unique, probe of the early
dust population.

The shift in central wavelength is particularly intriguing compared with the few UV bumps currently known during the reionization era. The feature in JADES-GS-z6-0 at $z=6.71$ peaks at
$\lambda_{\rm max}\simeq2263\,\AA$ and has a width of only $\sim250\,\AA$ \citep{Witstok2023,Lin2025}, while the $z=7.11$ galaxy studied by \citet{Ormerod2025} shows a central wavelength similarly redshifted, 
$\lambda_{\rm max}=2257^{+26}_{-28}\,\AA$. The previously identified
reionization-era bumps therefore peak redward of the Galactic value and have comparatively narrow profiles, whereas the feature reported here shifts in the opposite direction while retaining a Milky-Way-like width. The difference between our $\lambda_{\rm max}$ and those measured at $z\sim7$ is
$\simeq127\,\AA$, substantially larger than the quoted statistical uncertainties.

This comparison suggests that the diversity of UV-bump profiles
may itself contain important information on early dust
populations. If the features are produced by dust, their different
central wavelengths and widths argue against a universal
UV-bump profile at early cosmic times and suggest substantial
variations in the composition and/or processing state of the
carriers among young galaxies. In this sense, the shape of the
bump, rather than its detection alone, provides an additional
observable with which to investigate the properties of dust
during the first stages of galaxy evolution.

An alternative interpretation that deserves consideration is
whether the apparent continuum depression could instead be
related to an AGN contribution. The Rest-frame UV spectra of AGN
and quasars can contain a dense forest of mixed \feii{} and
\feiii{} emission lines, producing an iron pseudocontinuum that
can substantially modify the apparent shape of the underlying
UV continuum \citep[e.g.,][]{Vestergaard2001,Baldwin2004}. In principle, a sufficiently strong iron
pseudocontinuum could  bias the reconstruction of
the intrinsic continuum and mimic or modify a broad attenuation
feature.

Such an interpretation appears to be disfavored for our source. The
rest-frame optical line-ratio diagnostics presented in Sect.\,\ref{sec:agn}
are consistent with a possible composite star-forming plus AGN
interpretation, but provide no evidence that the source is
dominated by an extreme, highly ionized AGN. In particular,
the diagnostics involving the temperature-sensitive
\oiii\,$\lambda4363$ line place the source outside the regime
occupied by the most extreme AGN-dominated systems. 
An AGN contribution, potentially even substantial, cannot be robustly excluded with the available emission-line diagnostics, given the overlap between stellar and AGN photoionization models at these redshifts. Nevertheless, the absence of independent evidence for a luminous AGN, together with the observed Drude-like profile, leads us to favor dust attenuation over an AGN-related iron pseudocontinuum as the origin of the broad UV depression.

\subsection{What is the carrier of the $z=9.3335$ UV bump?}
The location of the $z=9.3335$ bump in the
$\lambda_{\max}$--FWHM plane provides an interesting constraint on its
possible carrier. Figure\,\ref{fig:bump_profile_space} shows that the feature
is displaced toward shorter central wavelengths than the bulk of the Milky Way
sightlines, whereas its width lies well within the Galactic distribution.
The distinction from the local population is therefore primarily driven by 
$\lambda_{\max}$.

The comparison with theoretical calculations illustrates how changes
in the carrier population can affect both observables. Uniformly
weighted PAH mixtures \citep{Lin2023} span widths broader than
our measurement over much of the range of PAH sizes explored by
those models, while their central wavelengths remain strongly
dependent on the adopted molecular mixture. The spherical
graphite-grain calculations of \citet{Li2024} approach the location
of our source in the $\lambda_{\rm max}$--FWHM plane for the
smallest grains, whereas increasing grain size shifts the feature
toward longer wavelengths and larger widths. Thus, unlike the
redshifted and relatively narrow bump observed in JADES-GS-z6-0,
for which spherical graphite grains provide a poor match, the
profile measured here does not clearly exclude a small-graphite-grain
origin.

The quantum-chemical calculations of individual PAH species by
\citet{Lin2025} span a substantially broader region of the
$\lambda_{\rm max}$--FWHM plane than uniformly weighted PAH
mixtures. This broad spread demonstrates that changes in molecular
composition can substantially shift the peak wavelength while 
producing a range of feature widths, including values comparable
to those observed here (Fig.\,\ref{fig:bump_profile_space}). Small PAHs
or related aromatic species therefore also provide a plausible
interpretation of the observed profile. 
However, the abundance and survival of small PAHs at such an early epoch remain uncertain. Observations at lower redshifts indicate that the PAH-to-dust mass fraction decreases to below $1\%$ at metallicities $Z\lesssim0.3\,Z_\odot$ \citep{Aniano2020,Shivaei2024}, while the hard radiation field inferred for our source may further hinder the survival of the smallest molecules. Thus, although PAH-based models can reproduce the observed bump profile, their physical viability in this low-mass, metal-poor and highly ionized galaxy remains to be established \citep{Pedrini2026}.

Overall, these comparisons suggest that the displaced central
wavelength of our source may encode information on the composition and size
distribution of its carbonaceous dust population. Only $\sim500\,\mathrm{Myr}$ after
the Big Bang, the population responsible for the UV bump needs not yet resemble
the mixture established in the evolved Galactic ISM. Different formation
pathways, selective survival in the intense radiation field, or limited
subsequent ISM processing could all contribute to such a difference. The
diversity between our source, the Milky Way, and the currently known
$z\sim7$ systems therefore suggests that structured UV attenuation was
already present at Cosmic Dawn, whereas the population responsible for the bump
may have varied substantially from galaxy to galaxy.

\subsection{Implications for dust formation at Cosmic Dawn}
Irrespective of the exact carrier, the presence of such a
structured UV attenuation feature at $z=9.3335$ places important
constraints on the timescale of dust evolution. Statistical studies
generally find weaker UV bumps at higher redshift, consistent
with the early dust populations being dominated by relatively large grains
freshly condensed in supernova ejecta and having experienced limited
ISM processing \citep{Markov2025}. The detection reported here shows
that such an average evolutionary trend does not preclude individual
galaxies from developing a dust population capable of producing a
pronounced UV feature at much earlier times. Models of dust evolution
during the first billion years predict an initially supernova-dominated
dust population followed by rapid grain growth and an increasing
small-to-large grain ratio as ISM processing becomes effective
\citep{Narayanan2025,Narayanan2026}.

The displacement of the bump central wavelength, despite a width
consistent with the Galactic distribution, may therefore be as
important as its early appearance. Rather than indicating that a
Milky-Way-like dust mixture was already established by $z=9.3335$,
our observations may be witnessing an earlier stage in the assembly
and processing of carbonaceous dust, in which the carrier population
differs from both present-day grains and those inferred in the
$z\sim7$ systems.

The presence of potential carbonaceous carriers at such early epochs
raises the question of how rapidly carbon-rich dust can form.
Core-collapse supernovae provide a possible early production channel,
although the efficiency of carbon-grain condensation depends on the
competition with CO molecule formation and on the thermal and chemical
evolution of the ejecta. Models exploring these processes
\citep{Todini2001,Chiaki2026} illustrate the uncertainties associated
with carbon-dust production in supernovae, including the effects of
ejecta cooling and radioactive heating. Chemical-evolution modeling
by \citet{Nanni2025} for JADES-GS-z6-0 suggests that AGB stars alone
cannot account for the inferred abundance of carbonaceous grains
associated with its UV bump, favoring efficient grain growth in the
ISM and/or substantial carbon-dust production by core-collapse
supernovae.

Importantly, the relevant timescale for dust enrichment is not the
full cosmic age of $\sim500\,\mathrm{Myr}$ at $z=9.3335$, but the
interval since the onset of star formation. For an illustrative
formation redshift of $z\sim14$, approximately
$200\,\mathrm{Myr}$ would have been available for enrichment.
Such a timescale could allow contributions from the most rapidly
evolving AGB stars, alongside core-collapse supernovae and
subsequent grain growth in the ISM
\citep{Valiante2009,SchneiderMaiolino2024}.
However, the efficiency and composition of dust production
depend on the stellar initial mass function, metallicity,
and star-formation history
\citep{Valiante2009,SchneiderMaiolino2024}.
The relative importance of these channels remains uncertain
and requires detailed theoretical modeling beyond the scope
of the present observational study.

More broadly, the diversity already emerging among the few UV bumps
known at these epochs suggests that their detailed profiles can trace
different stages of early dust evolution. Measurements of the central
wavelength, width, and strength across larger spectroscopic samples
may therefore provide a way to distinguish systematic evolution in
dust composition and processing from galaxy-to-galaxy variations at
Cosmic Dawn, where direct constraints from dust thermal emission
remain extremely challenging.


\section{Summary}

We have presented the identification and characterization of a UV
attenuation bump in the Black Pearl galaxy at $z_{\rm spec}=9.3335$, observed with JWST/NIRSpec as part
of the ZENITH+EMBER survey. This represents the highest-redshift UV bump
reported to date, probing dust properties only $\sim 500\,\mathrm{Myr}$ after
the Big Bang. Our main results can be summarized as follows.

\begin{itemize}

\item The broad depression observed in the rest-frame UV spectrum is
well described by a Drude profile, with $\lambda_{\rm max}=2132\pm17\,\AA$,
${\rm FWHM}=413\pm54\,\AA$, and a peak attenuation
$A_{\rm max}=0.21\pm0.02$ mag. The recovery of the continuum blueward
of the feature cannot be reproduced by a simple extrapolation of the
red-side power-law continuum.

\item Joint spectrophotometric modeling with a flexible attenuation
law independently recovers the UV bump, without imposing its presence
a priori. The feature is robust against the adopted star-formation
history, strengthening its interpretation as a genuine dust
attenuation feature.

\item Rest-frame optical emission-line diagnostics indicate a hard
ionizing spectrum and are consistent with intense star formation
possibly accompanied by an AGN contribution. However, the source does
not occupy the regime of the most extreme AGN-dominated systems,
making a strong AGN iron pseudocontinuum an unlikely explanation for
the observed broad UV feature.

\item The measured bump differs from both the canonical Milky Way
$2175\,\AA$ feature and the UV bumps currently known at $z\sim7$ primarily
in its central wavelength. Its FWHM is consistent with the Galactic
distribution, whereas its peak is shifted blueward; the known $z\sim7$
features instead peak redward and are comparatively narrow. This diversity
argues against a universal UV-bump profile at early cosmic times and suggests
that the carriers may already span different compositions and/or processing
states during the first stages of galaxy evolution.

\end{itemize}

Our results highlight the potential of UV-bump profiles as probes of dust composition and processing at Cosmic Dawn, providing insights into grain properties at epochs when direct constraints from dust thermal emission remain extremely challenging.

\begin{acknowledgments}
This work is supported  by the European Union -- NextGeneration EU RFF M4C2 1.1 PRIN 2022 project 2022ZSL4BL INSIGHT (P.I. B. Vulcani)
and by INAF RF2024 Large Grant "UNDUST: UNveiling the Dawn of the Universe with JWST" (P.I. P. Santini).
\end{acknowledgments}





%
\facilities{HST, JWST.}

\software{astropy \citep{2013A&A...558A..33A,2018AJ....156..123A,2022ApJ...935..167A},  
          Cloudy \citep{2013RMxAA..49..137F}, 
          Source Extractor \citep{1996A&AS..117..393B}
          }



\bibliography{sample701}{}
\bibliographystyle{aasjournalv7}

\appendix
\section{Comparison of SED fitting with and without the UV bump}
\label{sed-nobump-sibump}

\begin{figure}[h]
\centering
\includegraphics[width=9cm]{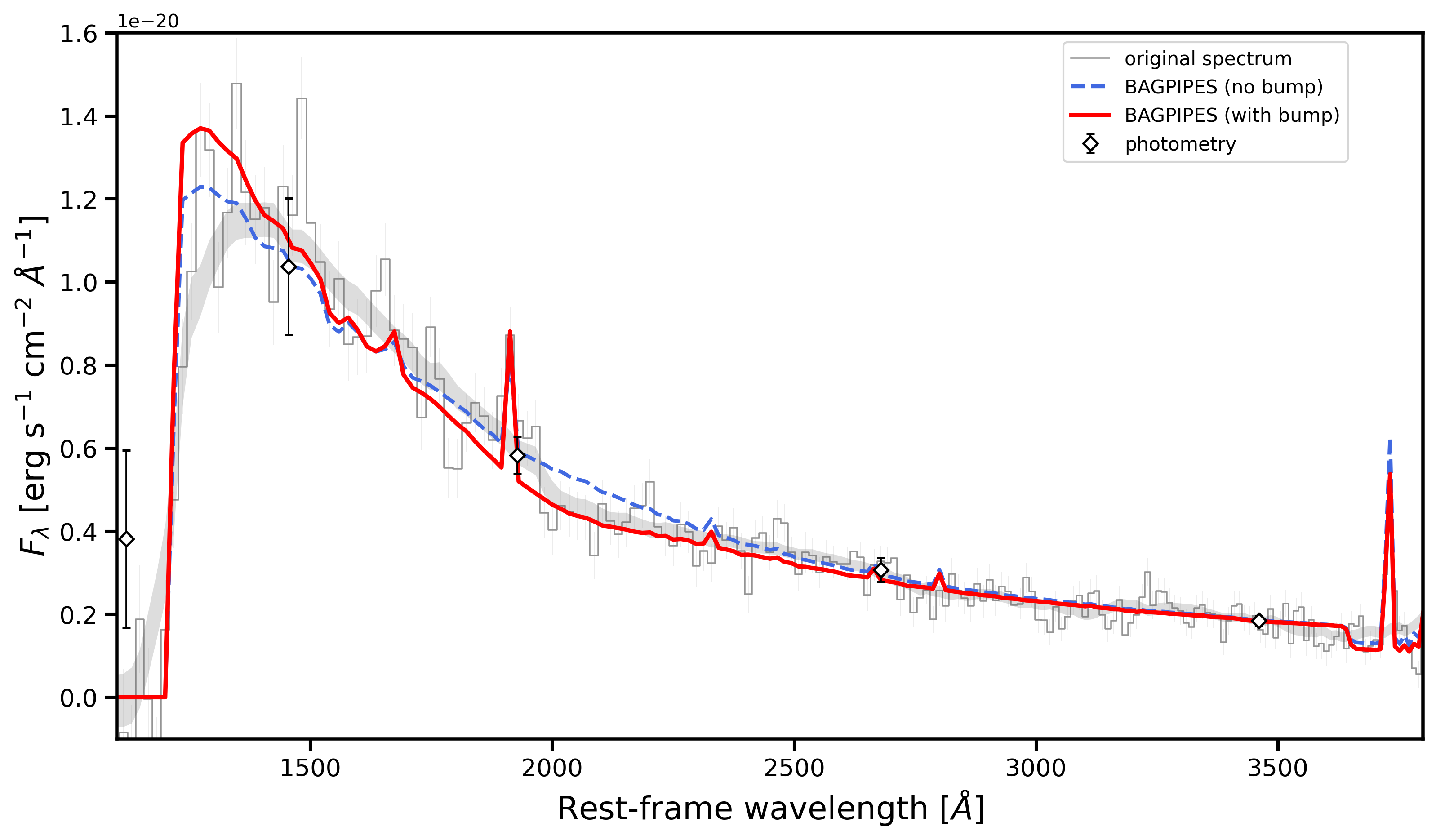}
\caption{Comparison of the \texttt{BAGPIPES} best-fitting SEDs for the Black Pearl, adopting the reference continuity star formation history and flexible dust attenuation prescription described in the main text. The red solid curve shows the model including the Drude UV-bump component, while the blue dashed curve corresponds to the fit with this component disabled. The observed NIRSpec spectrum is shown in gray, together with its smoothed representation and uncertainty envelope, while the black diamonds indicate the broadband photometry. Both models are plotted in the rest frame.}
\label{fig:fitfit}
\end{figure}

In Fig. \ref{fig:fitfit}, we compare the \texttt{BAGPIPES} best-fitting spectral energy distributions obtained with and without the UV attenuation bump for the Black Pearl. Both fits adopt the reference continuity star formation history and the flexible dust attenuation prescription described in Section \ref{sed_fitting}, differing only in whether the Drude component is enabled or disabled. The model including the bump provides a different description of the rest-frame UV continuum, while the two solutions become increasingly similar at longer wavelengths. The physical parameters inferred from the reference continuity-SFH fit are reported in Table \ref{tab:posterior}.

\section{Additional constraints on the ionizing source and nebular geometry}
\label{app:ionization}

Given the high \oiii/\oii{} ratio and the highly ionized
conditions inferred for the Black Pearl, we further
investigate the nature of the ionizing source and the geometry
of the ionized gas. The conventional emission-line diagnostics
presented in Sect.\,3.3 (Fig.\,6) do not provide an unambiguous
distinction between stellar and AGN photoionization.

To explore this issue, we follow the approach of
\citet{Borja2026} and examine the cumulative emission-line
contributions as a function of depth within a photoionized
cloud. This allows us to investigate whether the observed
line ratios can be reproduced by matter-bounded models,
in which the ionized region is truncated before reaching
the ionization front. We adopt a normalized cloud-depth
coordinate, $D_{\rm norm}$, defined such that
$D_{\rm norm}=1$ corresponds to the illuminated face
of the cloud.
The photoionization models are computed at fixed gas-phase
metallicity, $12+\log({\rm O/H})=7.99$, and ionization
parameter, $\log U \simeq -1.5$, motivated by the
SED-fitting results presented in Sect.~\ref{sed_fitting}. The ionization
parameter is held fixed for each assumed gas density.
We note, however, that the SED-derived ionization
parameter approaches the upper boundary of the adopted
prior and should therefore be regarded as indicative
rather than tightly constrained.

\begin{figure*}[h]
\centering
\includegraphics[width=15cm]{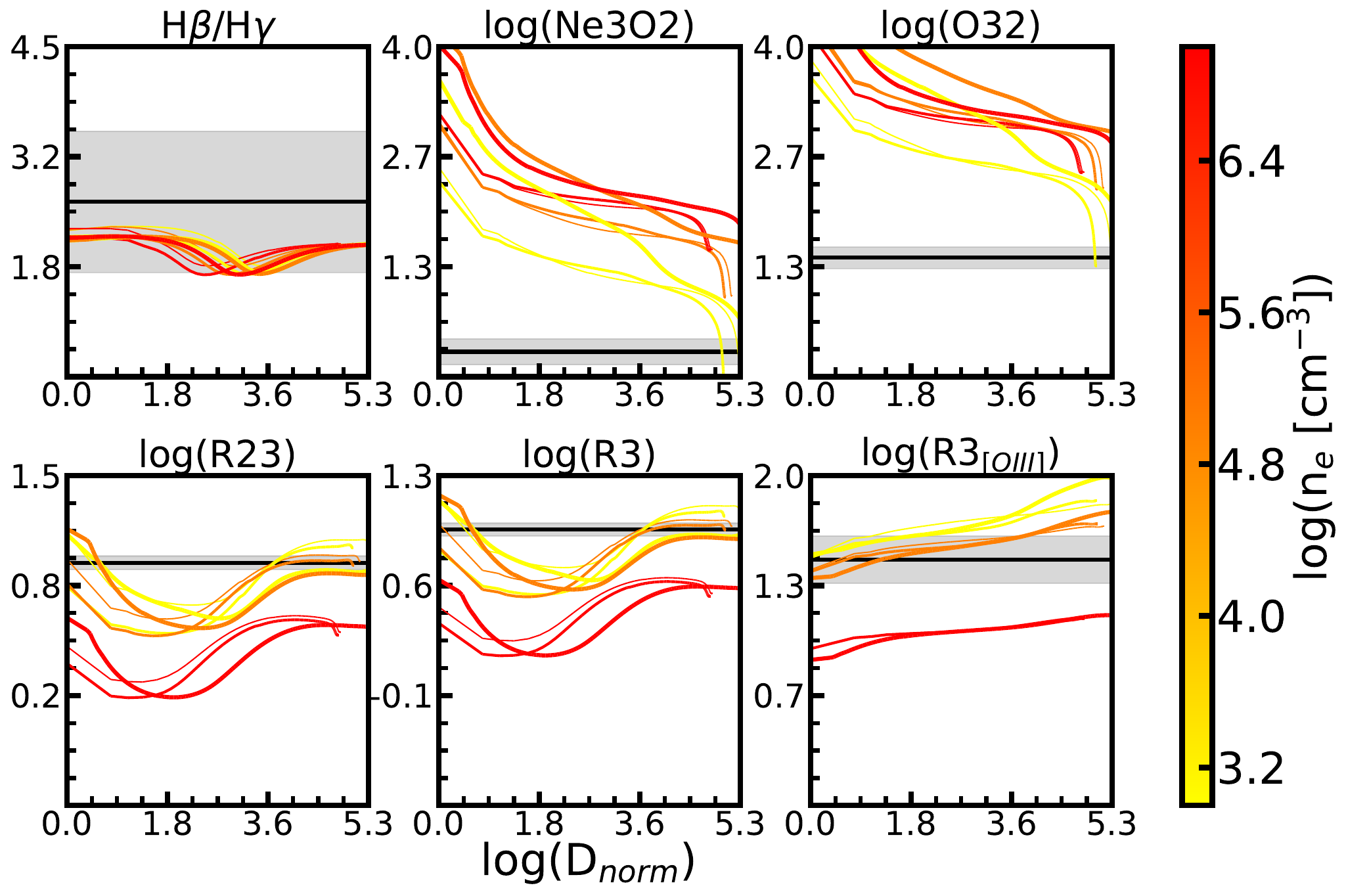}
\caption{
Predicted emission-line diagnostics as a function of
normalized cloud depth for the photoionization models
described in Appendix~\ref{app:ionization}.
The models assume a fixed gas-phase metallicity of
$12+\log({\rm O/H})=7.99$ and an ionization parameter
of $\log U\simeq-1.5$.
The comparison with the observed line ratios is used
to investigate the consistency of the data with
stellar photoionization under matter-bounded conditions.
}
\label{fig:borja}
\end{figure*}

Figure~\ref{fig:borja} presents the
predictions of these models. The diagnostic ratios
constructed from the stronger emission lines detected
in the NIRSpec/PRISM spectrum, including
H$\beta$/H$\gamma$, O32, Ne3O2, and R23, are broadly
consistent with photoionization by massive stars.
Within the adopted model framework, reproducing the
observed ratios favors a matter-bounded configuration,
in which the ionized region is truncated and a fraction
of the ionizing photons may escape.
Diagnostics involving the auroral \oiii\,$\lambda4363$
line, particularly O3Hg and the ratio of the strong
\oiii\,$\lambda\lambda4959,5007$ lines to
\oiii\,$\lambda4363$, appear to favor a more strongly
truncated ionized region. However, the
\oiii\,$\lambda4363$ detection has a significance
of only $\sim3\sigma$, resulting in substantial
uncertainties in these ratios. We therefore refrain
from drawing firm conclusions from the auroral-line
diagnostics alone, particularly given the more
consistent picture provided by the stronger emission
lines.

An independent, albeit indirect, indication of a
nonzero escape fraction can be obtained from the
observed UV continuum slope, $\beta=-2.22$.
Applying the empirical relation of
\citet{Chisholm2022}, we estimate an ionizing-photon
escape fraction of order
$f_{\rm esc}\sim0.03$--$0.10$, corresponding to
a characteristic value of approximately $6\%$.
This estimate is subject to the assumptions and
intrinsic scatter of the adopted empirical relation
and should not be interpreted as a direct measurement
of $f_{\rm esc}$.

Overall, the observed emission-line ratios are
compatible with photoionization by massive stars
in a moderately matter-bounded nebular configuration,
potentially allowing a modest escape of ionizing
photons. 
Although an AGN contribution cannot be
excluded, particularly in connection with the
\oiii\,$\lambda4363$ emission, the available
diagnostics do not require an AGN-dominated ionizing
spectrum. However, the existence of a viable stellar-photoionization solution does not rule out an AGN-dominated alternative, which would require joint modeling of stellar and AGN contributions to assess quantitatively.

\end{document}